\documentclass[prd,aps,amsfonts,longbibliography,notitlepage,nofootinbib,superscriptaddress,eqsecnum]{revtex4-1}
\usepackage{booktabs}
\usepackage{graphicx}
\usepackage[dvipsnames]{xcolor}
\usepackage{amsmath,amssymb,graphics,amsthm,isomath}
\usepackage{bbm}
\usepackage{bm}
\usepackage{array}
\usepackage{tabularx}
\newcolumntype{P}[1]{>{\centering\arraybackslash}p{#1}}
\usepackage{multirow}
\usepackage{subcaption}
\usepackage{comment}
\usepackage{soul}

\usepackage[colorlinks=true, urlcolor=violet, linkcolor=blue, citecolor=red, hyperindex=true, linktocpage=true]{hyperref}

\allowdisplaybreaks

\newcommand{\R}{\mathbb{R}}

\numberwithin{thm}{section}

\makeatletter
\renewcommand{\p@subsection}{}
\renewcommand{\p@subsubsection}{}
\makeatother

\renewcommand{\arraystretch}{1.4}

\newcommand{\andy}[1]{{\color{red} Andy:~#1}}

\begin{document}
\title{Quadrupole-conserving dynamics in the non-commutative plane}

\author{Isabella Zane}
\affiliation{Department of Physics and Center for Theory of Quantum Matter, University of Colorado, Boulder, CO 80309, USA}

\author{Andrew Lucas}
\email{andrew.j.lucas@colorado.edu}
\affiliation{Department of Physics and Center for Theory of Quantum Matter, University of Colorado, Boulder, CO 80309, USA}

\begin{abstract}
Inspired by ``fracton hydrodynamic" universality classes of dynamics with unusual conservation laws, we present a new dynamical universality class that arises out of local area-preserving dynamics in the non-commutative plane.  On this symplectic manifold, the area-preserving spatial symmetry group $\mathrm{SL}(2,\mathbb{R})\rtimes \mathbb{R}^2$ is a symmetry group compatible with non-trivial many-body dynamics.  The conservation laws associated to this symmetry group correspond to the dipole and quadrupole moments of the particles.  We study the unusual dynamics of a crystal lattice subject to such symmetries, and argue that the hydrodynamic description of lattice dynamics breaks down due to relevant nonlinearities.  Numerical simulations of classical Hamiltonian dynamical systems with this symmetry are largely consistent with a tree-level effective field theory estimate for the endpoint of this instability.
\end{abstract}

\date{\today}
\maketitle

\tableofcontents

\section{Introduction}
The past decade has seen a number of works developing new phases of matter with fracton excitations \cite{Nandkishore:2018sel,Pretko:2020cko,RevModPhys.96.011001}.   Here, we refer to a ``fracton" as an excitation which has restricted mobility, which is often relaxed when it can move in tandem with another fracton.

In many of the simplest cases, the immobility of single excitations can be captured by the presence of an unusual U(1) symmetry, i.e. conservation law.   As a simple example, consider a theory of particles interacting on a line, but where the center of mass $\sum x_i$ of every particle is a conserved quantity.\footnote{This is a conserved quantity \emph{in the center of mass frame} of a Galilean-invariant theory.  In a fractonic theory, one would demand instead that this quantity is conserved in every reference frame \cite{Glorioso:2021bif}.}  A single particle cannot move on its own, but two particles can move towards or apart from each other.  It is standard in the literature to think of each particle as carrying a charge, and to say that this model has dipole conservation.  

Previous works \cite{gromov2018towards,Bulmash:2023msp} have classified many of the possible theories with such multipolar conservation laws.  For simplicity containing the discussion to one spatial dimension, a symmetry algebra of particular interest is called the \emph{multipole algebra}, where the conserved quantities are the momentum $P$ and the multipole moments $X_0,\ldots, X_m$ where \begin{equation}
    X_k = \sum_{a=1}^N x_a^k.
\end{equation}
Here $x_a$ represents the position of each particle on the line.  Using the canonical Poisson brackets $\lbrace x_a, p_b\rbrace = \delta_{ab}$, one finds that the multipole algebra is closed: \begin{equation}
    \lbrace X_k, P\rbrace = kX_{k-1}. \label{eq:intro_multipole}
\end{equation}Explicit Hamiltonian models with this symmetry group were presented in \cite{osborne}.

One particularly interesting thing about theories with a multipolar symmetry group is that the hydrodynamic relaxation of such theories to thermal equilibrium generally lies in a distinct universality class \cite{osborne,gromovfrachydro,knap2020,hart2021hidden,Grosvenor:2021rrt,Glodkowski:2022xje,Glorioso:2021bif,Glorioso:2023chm,Guo:2022ixk,Jain:2023nbf,Glodkowski:2024ova}, compared to standard theories of liquids and gases.   These so-called ``fracton hydrodynamic" theories have clear imprints of the multipole symmetry in the dispersion relation of sound waves, which would scale as $\omega \sim k^{m+1}$ when $X_0,\ldots, X_m$ are all conserved quantities \cite{osborne}.   While the experimental applicability of many of these theories is unclear, outside of e.g. realizing dipole-conserving subdiffusion in a tilted optical lattice \cite{guardado2020subdiffusion}, it is nevertheless worthwhile to understand the space of possible hydrodynamic theories consistent with the laws of statistical mechanics.

In this paper, we will look for hydrodynamic universality class that lies ``beyond" a theory with a symmetry group like \eqref{eq:intro_multipole}. One example which has already appeared in the literature are modulated symmetries \cite{sala2021dynamics}; another example are subsystem symmetries \cite{Iaconis:2019hab}.  What appears to us to be the simplest example which does not quite fall into these existing categories is a model of particles in the non-commutative plane $\mathbb{R}^2$ with dipole and quadrupole conservation.  Dynamics in this non-commutative space are relevant for the dynamics of vortices in two-dimensional fluids or superfluids \cite{doshi}, as well as to particles in the lowest Landau level in a quantum Hall regime \cite{gromovfrachydro,DuMehtaSon2024}, as two examples.  A phase space interpretation of the same algebra was presented in \cite{Sadki:2025bll}.   While at first glance, this seems like a regular multipole algebra, the non-commutative coordinates render the algebra different from \eqref{eq:intro_multipole}.  In fact, we will argue that this theory of quadrupole-conserving dynamics is as symmetric as possible, without being trivial.   

We now present a roadmap to the rest of the paper.  In Section \ref{sec:symmetry} we discuss the symmetry group of interest and Hamiltonian mechanics in the non-commutative plane.  Section \ref{sec:hamiltoniansym} describes how to build symmetric Hamiltonians, which turn out to have an elegant interpretation: they are the areas of polygons that triangulate the plane.  We therefore deduce that all symmetric models have at least three-body interactions.   Section \ref{sec:lattices} analyzes some simple examples of many-body systems with this symmetry group, where we find models with many accidental symmetries at low energy, and a general dispersion relation $\omega \sim k^3$ for sound waves.   A general effective field theory for non-dissipative dynamics is developed in Section \ref{sec:eftnondiss}, with the dissipative generalization presented in Section \ref{sec:eftdiss}.  We predict that the hydrodynamic description of sound waves breaks down due to relevant nonlinearities in fluctating hydrodynamics.  Numerical simulations, presented in Section \ref{sec:numerics}, confirm this expectation.

\section{Symmetry}\label{sec:symmetry}
In this paper, we begin by considering classical Hamiltonian dynamics on phase space $\mathbb{R}^{2N}$, with canonical (Darboux) symplectic form \begin{equation}
    \omega = \sum_{a=1}^N \mathrm{d}y_a \wedge \mathrm{d}x_a.
\end{equation}
The Poisson brackets are \begin{equation}
    \lbrace x_a,y_b\rbrace = \delta_{ab}.
\end{equation}
In physical terms, we can think of our phase space as describing $N$ particles in a lowest Landau level \cite{DuMehtaSon2024}, or $N$ quantized clockwise-rotating vortices in a superfluid thin film \cite{doshi}.   In what follows, importantly, we will consider the dynamics as taking place in two-dimensional space; the corresponding notions of two-dimensional locality will play an essential role in everything that follows.

Note that in this paper, we will always use $ab$ indices to denote particles, while $x_{a,i}=(x_a,y_a)$ will be used to denote the two-component position vector of a single particle, with $ij$ indices used to denote spatial directions.

\begin{figure}
\centering
\begin{tabular}{c c c c}
\hline
Function & Vector Field & Illustration & Interpretation \\
\hline

 $N = \sum_a 1$& $0$ 
&  & trivial \\[1em]

 $X = \sum_a x_a$& $-\sum_a \frac{\partial}{\partial y_a}$ 
& \includegraphics[keepaspectratio,width=1cm]{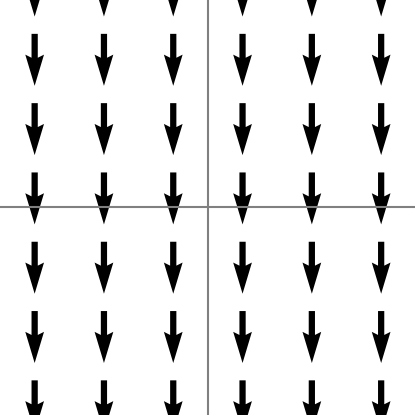} & translation (down) \\[1em]

$Y = \sum_a y_a$ & $\sum_a \frac{\partial}{\partial x_a}$ 
& \includegraphics[keepaspectratio,width=1cm]{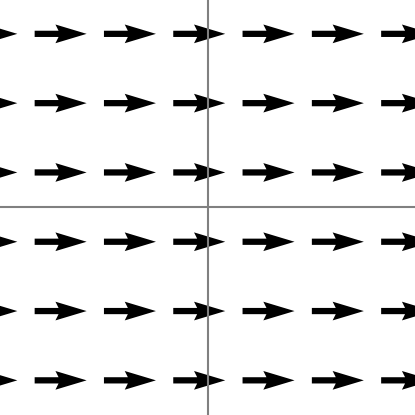} & translation (right) \\[1em]

$L =-\frac{1}{2} \sum_a (x_a^2 +y_a^2)$ & $\sum_a \bigl(x_a\partial_{y_a}-y_a\partial_{x_a}\bigr)$ 
& \includegraphics[keepaspectratio,width=1cm]{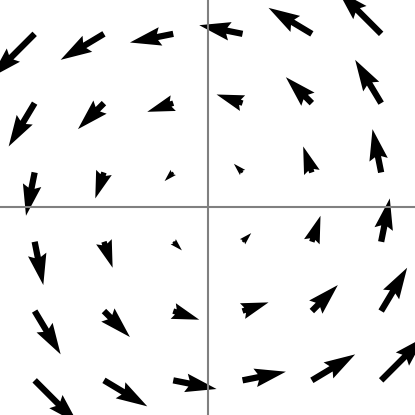} & rotation \\[1em]

$Q = \frac{1}{2}\sum_a (x_a^2 - y_a^2)$ & $-\sum_a \bigl(x_a\partial_{y_a}+y_a\partial_{x_a}\bigr)$ 
& \includegraphics[keepaspectratio,width=1cm]{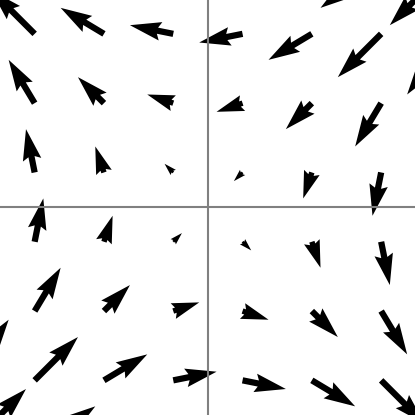} & hyperbolic shear \\[1em]

$R = \sum_a x_ay_a$ & $\sum_a \bigl(x_a\partial_{x_a}-y_a\partial_{y_a}\bigr)$ 
& \includegraphics[keepaspectratio,width=1cm]{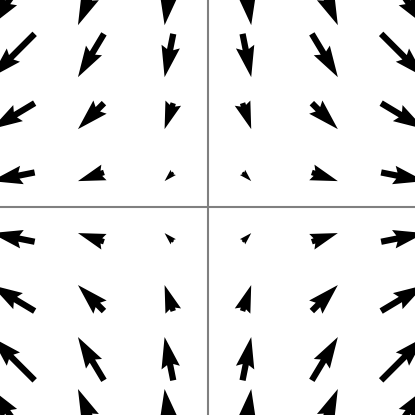} & squeeze \\[1em]
\hline
\end{tabular}
\caption{Summary of conserved quantities, their Hamiltonian vector fields, and geometric interpretations.}
\label{fig:symmetrygenerators}
\end{figure}

We consider Hamiltonian dynamical systems with a symmetry group $G$ consisting of canonical transformations (symplectomorphisms) generated by the following six elements: \begin{equation} \label{eq:symmetrygenerators}
    \left(\begin{array}{c} N \\ X \\ Y \\ L \\ Q \\ R \end{array}\right) = \sum_{a=1}^N \left(\begin{array}{c} 1 \\ x_a \\ y_a \\ -\frac{1}{2}(x_a^2+y_a^2) \\ \frac{1}{2}(x_a^2-y_a^2) \\ x_ay_a \end{array}\right).
\end{equation}
An intuitive understanding of each of these symmetry generators is presented in Figure \ref{fig:symmetrygenerators}.  
\setlength{\tabcolsep}{12pt}
In more detail, let us analyze what this symmetry amounts to.  It is straightforward to check that these 6 conserved quantities span a closed Lie algebra, whose non-trivial commutators (Poisson brackets) are: 
%\andy{add the rest!} 
\begin{subequations}
    \begin{align}
        \lbrace X,Y\rbrace &= N \\
        \lbrace L, Y\rbrace &= -X \\
        \lbrace X, L\rbrace &= -Y \\
        \lbrace X, Q\rbrace &= -Y \\
        \lbrace X, R\rbrace &= X \\
        \lbrace Y,Q\rbrace &= -X \\
        \lbrace Y,R \rbrace &= -Y\\
        \lbrace L,Q\rbrace &= 2R\\
        \lbrace L,R\rbrace &= -2Q\\
        \lbrace Q,R\rbrace &= -2L
    \end{align}
\end{subequations}
As in \eqref{eq:intro_multipole}, we can understand this as a kind of quadrupolar algebra on the non-commutative plane.  We will shortly justify that this algebra is very special.  

Let us deduce the canonical transformations generated by these 6 conserved quantities.  Recall that the symplectic form allows us to uniquely associate every function $f$ with a vector field $X_f$ (a standard but, for us, unfortunate notation) by:
\begin{equation}
    \mathrm{d}f(Y) = \omega(X_f, Y)
\end{equation}
Each conserved quantity generates a Hamiltonian vector field: \begin{subequations}
    \begin{align}
        X_X &= -\sum_{a=1}^n \frac{\partial}{\partial y_a}, \\
        X_Y &= \sum_{a=1}^n \frac{\partial}{\partial x_a}, \\
        X_L &= \sum_{a=1}^n \left(x_a \frac{\partial}{\partial y_a} - y_a\frac{\partial}{\partial x_a}\right), \\
        X_Q &= -\sum_{a=1}^n \left(x_a\frac{\partial}{\partial y_a} + y_a\frac{\partial }{\partial x_a}\right), \\
        X_R &= \sum_{a=1}^n \left(x_a \frac{\partial}{\partial x_a} -y_a\frac{\partial }{\partial y_a}\right)
    \end{align}
\end{subequations}
with $X_N=0$.  The canonical transformations generated by these transformations are quite simple: \begin{subequations}
    \begin{align}
        \mathrm{e}^{tX_X} \left(\begin{array}{c} x_a \\ y_a \end{array}\right) &= \left(\begin{array}{c} x_a \\ y_a - t \end{array}\right), \\
        \mathrm{e}^{tX_Y} \left(\begin{array}{c} x_a \\ y_a \end{array}\right) &= \left(\begin{array}{c} x_a +t \\ y_a \end{array}\right), \\
        \mathrm{e}^{tX_L} \left(\begin{array}{c} x_a \\ y_a \end{array}\right) &= \left(\begin{array}{cc} \cos t &\ -\sin t \\ \sin t &\ \cos t \end{array}\right)\left(\begin{array}{c} x_a \\ y_a \end{array}\right), \\ 
        \mathrm{e}^{tX_Q} \left(\begin{array}{c} x_a \\ y_a \end{array}\right) &= \left(\begin{array}{cc} \cosh t &\ -\sinh t \\ -\sinh t &\ \cosh t \end{array}\right)\left(\begin{array}{c} x_a \\ y_a \end{array}\right), \\
        \mathrm{e}^{tX_R} \left(\begin{array}{c} x_a \\ y_a \end{array}\right) &= \left(\begin{array}{cc} \mathrm{e}^{-t} &\ 0 \\ 0 &\ \mathrm{e}^t \end{array}\right)\left(\begin{array}{c} x_a \\ y_a \end{array}\right).
    \end{align}
\end{subequations}
So $X$ and $Y$ generate translations, and represent momenta in the non-commutative plane.   $L$, $Q$ and $R$ generate linear transformations of the plane $\mathbb{R}^2$, and it is easily seen that each of them preserve area: the determinant of each $2\times 2$ matrix is 1.  The group of such area-preserving linear transformations is $\mathrm{SL}(2,\mathbb{R})$.  $X$ and $Y$ generate an affine extension of $\mathrm{SL}(2,\mathbb{R})$, such that the full symmetry group is $G = \mathrm{SL}(2,\mathbb{R})\rtimes \mathbb{R}^2$.  The symmetry algebra represents a central extension (by $N$) of the Lie algebra of $G$, in which translations are non-commutative in the plane.   
%We remark that this symmetry group is a generalization of the much more common Euclidean symmetry group $\mathrm{E}(2) = \mathrm{SO}(2)\rtimes \mathbb{R}^2$, corresponding to rotations and translations in the two-dimensional plane.

Let us now argue that this symmetry group $G$ is ``uniquely  interesting" in the non-commutative plane.  Without the generators $R$ and $Q$, this is the symmetry algebra of particles in the lowest Landau level \cite{DuMehtaSon2024}, or  interacting vortices in two-dimensional superfluid films \cite{doshi}, and the dynamics of such systems been studied previously.  Without $L$ as well, we find a non-commutative dipole algebra which has been studied recently \cite{DuMehtaSon2024} in the context of lowest Landau level physics.  Let's now ask how we might generalize this non-commutative dipole algebra in an interesting way.  If we start with $N$, $X$ and $Y$ alone, we can consider a sort of non-commutative multipole algebra in \emph{one} of the two dimensions: letting \begin{equation}
    X_k := \sum_{a=1}^n x_a^k
\end{equation} with $X_1=X$ and $X_0=N$, we see that \begin{equation}
    \lbrace X_k, Y\rbrace = k X_{k-1}.
\end{equation}
So we can consider a symmetry group generated by $X_k,X_{k-1},\ldots, X_0, Y$, which looks like a standard multipole algebra \cite{gromov2018towards}, as in \eqref{eq:intro_multipole}.  In the non-commutative plane, of course, $Y$ simply generates translations in the $x$-direction, so plays the role of momentum in the usual multipole algebra.

Our full algebra \eqref{eq:symmetrygenerators} is the unique symmetry algebra which is non-trivial, while also lying beyond the examples above, which have all been previously studied.  To see this, notice that if we consider \begin{equation}
    \lbrace x^n y^m, x^{n^\prime}y^{m^\prime}\rbrace = \left(nm^\prime - n^\prime m\right)x^{n+n^\prime-1}y^{m+m^\prime-1},
\end{equation}
clearly the algebra of higher-order polynomials will not close if $n+m>2$ is possible, unless $nm=0$ for every term in the algebra (as in the multipole case).  Therefore, if we have any $x^2$, $y^2$, or $xy$ in a symmetry generator, the only finite algebras are the vortex algebra or \eqref{eq:symmetrygenerators}.  If we have an infinite algebra of all polynomials, then for a finite number $n$ of particles, we expect there to be no dynamics (particle motion) that are compatible with the infinitely many conservation laws.

The reader may ask why we cannot invoke the \emph{infinite-dimensional} two-dimensional conformal symmetry group \cite{cft} as a kind of non-trivial multipolar symmetry group.   The reason this is not allowed is that these transformations are not canonical, and thus do not leave the symplectic form on phase space invariant.  This is most easily seen by converting to complex coordinates $z=x+\mathrm{i}y$ and $\bar z = x-\mathrm{i}y$; the symplectic form becomes \begin{equation}
    \omega = \frac{1}{2\mathrm{i}} \sum_{a=1}^N \mathrm{d}z_a \wedge \mathrm{d}\bar z_a.
\end{equation}
Under a conformal transformation $z\rightarrow f(z)$ with $f(z)$ an analytic function, and \begin{equation}
    \omega \rightarrow |f^\prime(z)|^2 \omega.
\end{equation}
Thus the subset of conformal transformations that are canonical correspond to \begin{equation}
    f(z) = \mathrm{e}^{\mathrm{i}\phi}z + a + \mathrm{i}b
\end{equation}
where $a,b,\phi \in \mathbb{R}$.  We see that the ``vortex subgroup" of \eqref{eq:symmetrygenerators} is precisely the intersection of conformal and canonical transformations. 

\section{Hamiltonian mechanics}\label{sec:hamiltoniansym}
Let us now describe how to find Hamiltonians which commute with all six generators in \eqref{eq:symmetrygenerators}.  

\subsection{Invariant building blocks}

We do so by finding ``invariant building blocks"  which are simple to physically interpret.  Translation symmetry tells us that the invariants can only depend on the difference between coordinates: $x_i-x_j$ or $y_i-y_j$. ``Modding out by translations" by working only with these differences, it remains to find $\mathrm{SL}(2,\R)$-invariant building blocks, keeping in mind that all $x$ and $y$ transform identically.  

Clearly, invariant building blocks of $\mathrm{SL}(2,\mathbb{R})$ must be invariant under each of the generators.  $L$ generates rotations, so if we collect $(x_a,y_a) \rightarrow x_{ia}$ where $i,j$ represent spatial indices, we must contract indices in a rotation-invariant way: namely, with the $\delta_{ij}$ and $\epsilon_{ij}$ tensors (or their tensor products).  $\delta_{ij}$ is invariant under neither $Q$ or $R$, while $\epsilon_{ij}$ is invariant under both.  Because $\epsilon_{ij}$ is the tensor used to define $\det M = M_{ij}M_{kl}\epsilon_{il}\epsilon_{jk}$, and determinants are invariant under $\mathrm{SL}(2,\mathbb{R})$, this conclusion is intuitive. We thus find our first invariant building block to be
\begin{equation}
    M(x_{1i},x_{2i},x_{3i},x_{4i}) = \epsilon_{ij}(x_{1i}-x_{2i})(x_{3i}-x_{4i}).
\end{equation}
And in fact, notice that \begin{equation}
    M(x_{1i},x_{2i},x_{3i},x_{4i}) = M(x_{1i},x_{2i},x_{1i},x_{4i}) - M(x_{1i},x_{2i},x_{1i},x_{3i}),
\end{equation}
and \begin{equation}
    M(x_{1i},x_{2i},x_{1i},x_{3i})= |(\vec{x}_3 - \vec{x}_1)\times (\vec{x_2}-\vec{x}_1)|= 2 \times [\text{Area of triangle with corners at $(\vec x_1,\vec x_2,\vec x_3)$}]. \label{eq:trianglearea}
\end{equation}
So the area of a triangle is an invariant building block for a Hamiltonian.  Note that in \eqref{eq:trianglearea} we are using the signed area of the triangle, which also keeps track of its orientation.

In fact, these are \emph{all} of the invariants.  After all, if we look for higher-order invariants, we will have to use $\epsilon$ to contract indices, and each index contraction must be with a difference of coordinates, which means that it will be a product of $M$s, and each $M$ in turn is a sum of fundamental building block \eqref{eq:trianglearea}.  

A more simple and physically transparent way to state our conclusion is that \emph{all} invariant functions under the symmetry group correspond to sums and products of the \emph{areas} of polygons in the plane, whose corners are formed by the particles at positions $(x_a,y_a)$.  For this area to not be zero identically, such a polygon needs at least three vertices.  Indeed, since lengths and angles are not preserved by shear transformations (generated by $R$ and $Q$), $\mathrm{SL}(2,\mathbb{R})$ has no nontrivial scalar invariants acting on one or two points. We conclude that all invariant theories that have non-constant Hamiltonians are necessarily interacting and have at least three-body interactions.  
%Consequently, the smallest nontrivial $\mathrm{SL}(2,\mathbb{R})$-invariant interactions necessarily involve at least three particles.
\begin{comment}
Putting it all together, the invariant building blocks of our symmetry group $G$ must be of the form
\begin{equation}
    A = (\vec{x}_1 - \vec{x}_2)_i (\vec{x}_1-\vec{x}_3)_j\epsilon_{ij}
\end{equation}

\andy{need to also point out that $x_{12}x_{34} = x_{12}x_{14}-x_{12}x_{13}$}
Notice that this is just twice the area of a triangle. 

\end{comment}

\subsection{Dynamics of a single triangle}
The simplest Hamiltonian we can write down involves 3 particles.  Let us now show that this Hamiltonian does generate dynamics, and can be solved in complete generality.  From the above discussion, the most general possible Hamiltonian is
\begin{equation}
    H = F\left(\epsilon_{ij}(x_{1i}-x_{2i})(x_{1j}-x_{3j}) \right) = F(A),
\end{equation}
where  below we use $A$ for simplicity to denote the area of the triangle formed by the 3 particles. Hamilton's equations of motion are
\begin{equation}
    \dot{x}_{ai} = \{x_{ai},H\}= \frac{\partial H}{\partial x_{aj}}\epsilon_{ij}.
\end{equation}
We find that
\begin{subequations}
\begin{align}
    \dot{x}_{1i} &= -(x_{2i} -x_{3i})F'(A) \\
    \dot{x}_{2i} &= -(x_{3i} -x_{1i})F'(A) \\
    \dot{x}_{3i} &= -(x_{1i} -x_{2i})F'(A) 
\end{align}
\end{subequations}
Since area is an invariant, $F'(A)$ is constant.  Therefore we can easily solve these equations in generality:
\begin{subequations}
\label{single_triangle_eq}
    \begin{align}
        x^1(t)&= \frac{1}{3} \left((2 c_1-c_2-c_3) \cos \left(\sqrt{3} F'(A) t\right)-\sqrt{3} (c_2-c_3) \sin \left(\sqrt{3} F'(A) t\right)+c_1+c_2+c_3\right) \\
        x^2(t) &= \frac{1}{3} \left(-(c_1-2 c_2+c_3) \cos \left(\sqrt{3} F'(A) t\right)+\sqrt{3} (c_1-c_3) \sin \left(\sqrt{3} F'(A) t\right)+c_1+c_2+c_3\right) \\
        x^3(t) &= \frac{1}{3} \left(-(c_1+c_2-2 c_3) \cos \left(\sqrt{3} F'(A) t\right)-\sqrt{3} (c_1-c_2) \sin \left(\sqrt{3} F'(A) t\right)+c_1+c_2+c_3\right) \\
         y^1(t)&= \frac{1}{3} \left((2 b_1-b_2-b_3) \cos \left(\sqrt{3} F'(A) t\right)-\sqrt{3} (b_2-b_3) \sin \left(\sqrt{3} F'(A) t\right)+b_1+b_2+b_3\right) \\
        y^2(t) &= \frac{1}{3} \left(-(b_1-2 b_2+b_3) \cos \left(\sqrt{3} F'(A) t\right)+\sqrt{3} (b_1-b_3) \sin \left(\sqrt{3} F'(A) t\right)+b_1+b_2+b_3\right) \\
        y^3(t) &= \frac{1}{3} \left(-(b_1+b_2-2 b_3) \cos \left(\sqrt{3} F'(A) t\right)-\sqrt{3} (b_1-b_2) \sin \left(\sqrt{3} F'(A) t\right)+b_1+b_2+b_3\right)
    \end{align}
\end{subequations}

Despite the rather cumbersome form of the equations themselves, they describe a very simple dynamical system, whereby the triangle formed by the three points moves along an ellipse.  A simple way to understand this system is to perform a canonical transformation such that the triangle is mapped to an equilateral triangle (of suitable area) whose center point is at the origin $(x,y)=0$.  Then $L$ is the only non-vanishing conserved quantity.  As all conserved quantities are invariant if the equilateral triangle rotates at a constant angular velocity, this is the resulting dynamics of the system.  Undoing these canonical transformations, we find that the general dynamics of the triangle corresponds to a ``precession" around a circumscribing ellipse.  A similar calculation was performed in \cite{Sadki:2025bll}.

\section{Dynamics of crystal lattices}\label{sec:lattices}
Now, we turn to a study of many-body dynamics.  In this section, we will focus on writing down simple many-body Hamiltonians that model crystals in the non-commutative plane (undergoing local area-preserving dynamics).   In the linear response regime, we will obtain the normal modes of these models exactly and discuss some curious features.

As shown in Section \ref{sec:hamiltoniansym}, the invariant motifs are the areas of shapes. A particularly simple Hamiltonian, which is bounded from below, is 
\begin{equation}
    H = \sum_{\text{polygon }s} \alpha_s (A_s-A_{s0})^2
\label{eq:Hamiltonian_triangle}
\end{equation}
where $A_s$ denotes the area of polygon $s$ and $\alpha_s \ge 0$ is some constant.  For simplicity in this section, we will always take $\alpha_s=1$ to be the same for all polygons.  For simplicity, we will also take $A_{s0}=A_0$ to be independent of polygon $s$.   In the examples that we will detail below, it will be possible to form a regular lattice where all $s$ indeed have the same area, meaning that the ground state $H=0$ can be achieved.  We will thus content ourselves with finding the normal modes of such Hamiltonians. 

\subsection{Triangular lattice}
\begin{figure}
    \centering
    \includegraphics[width=0.6\linewidth]{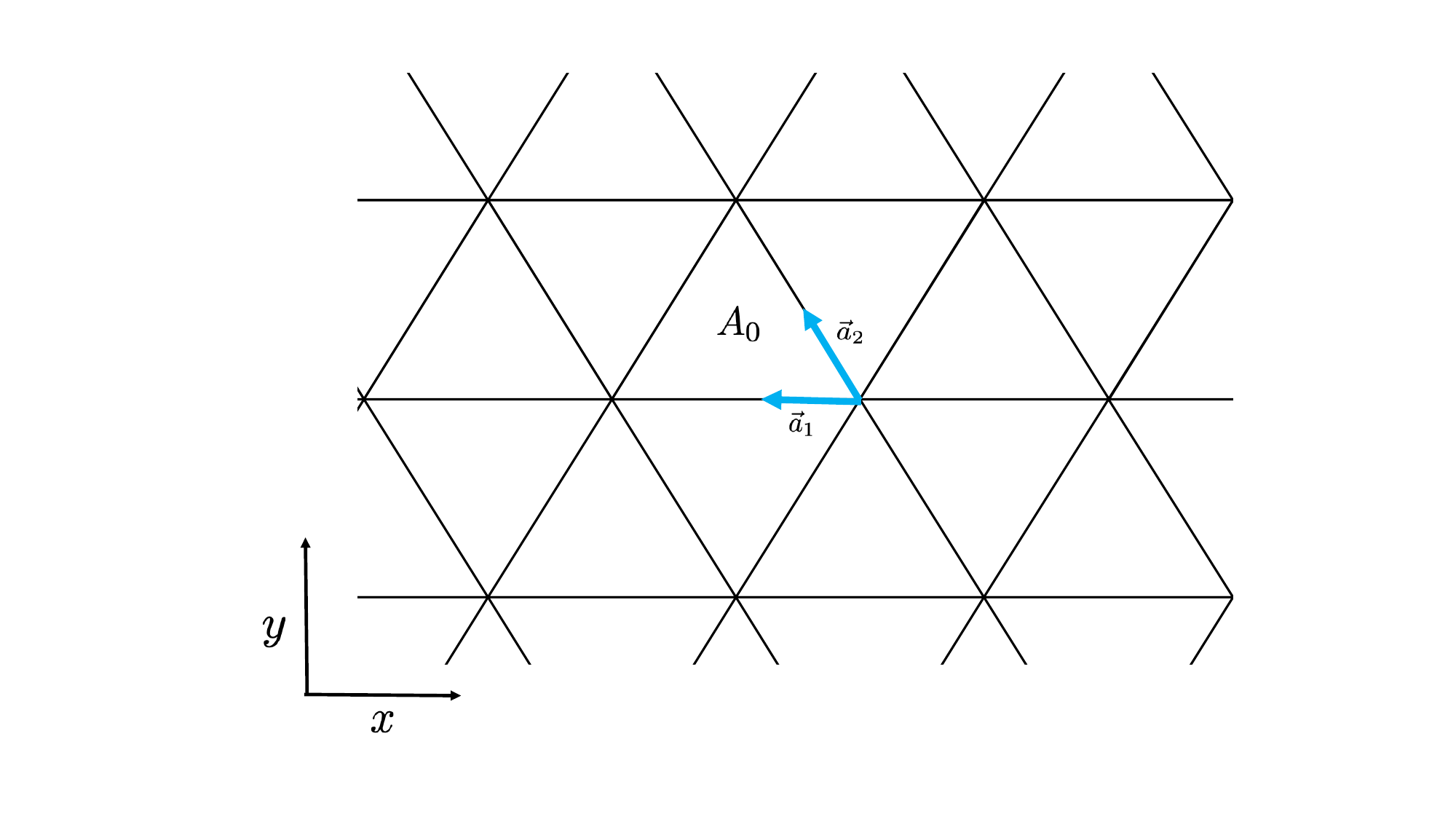}
    \caption{Triangular lattice with the ground state $A_s = A_0$ for all triangles $s$. Each vertex is a particle. }
    \label{fig:Perfect_triangular_lattice}
\end{figure}
Let us first discuss an illustrative example, where we have a triangular lattice, as depicted in Figure \ref{fig:Perfect_triangular_lattice}.  Given this lattice, which represents the ground state of \eqref{eq:Hamiltonian_triangle} so long as the polygons $s$ are chosen to be the equilaterial triangles making up the lattice, we may proceed to compute the normal modes. For small perturbations from equilibrium, the position of a particle located at site $(i,j)$ on the lattice is given by 
\begin{equation}
    \vec{x}_{ij} =  (\vec{x}_{\mathrm{eq}})_{ij} + \delta \vec{x}_{ij} \label{eq:xijpert}
\end{equation}
where the equilibrium positions of particles on the triangular lattice may be chosen to be \begin{equation}
    (\vec{x}_{\mathrm{eq}})_{ij} = \left(i + \frac{j}{2},\frac{\sqrt{3}j}{2}  \right) \label{eq:xeq_triangular}
\end{equation}
for integers $i$ and $j$.  
Recall that the area $A_s$ of each triangle is given by \eqref{eq:trianglearea}, upon using the perturbed coordinates \eqref{eq:xijpert}.

Given such a Hamiltonian, let us now outline a general formula for finding the normal modes. For small perturbations from equilibrium, the perturbed area is $A_0 + \delta A$, where we only keep linear order in $\delta \vec x$ when evaluating $\delta A$, as this is sufficient to keep all quadratic terms in $H \sim (\delta A)^2$. Let $(x_1,x_2,x_3)$ denote three particles that form a nearest-neighbor triangle, e.g $(x_{i,j},x_{i,j+1},x_{i+1,j})$. The change in area is given by  
\begin{comment}
\begin{equation}
    \delta A = \frac{1}{2}\left[(x^0_3-x^0_1)(\delta y_2 -\delta y_1) - (y^0_3-y^0_1)(\delta x_2 -\delta x_1) + (\delta x_3-\delta x_1)(y^0_2-y^0_1)-(\delta y_3-\delta y_1)(x^0_2-x^0_1)\right]
\end{equation}
\end{comment}
\begin{equation}
    \delta A = \frac{1}{2}\left[\mathrm{\Delta} x_{31}(\delta y_2 -\delta y_1) - \mathrm{\Delta} y_{31}(\delta x_2 -\delta x_1) + \mathrm{\Delta} y_{21}(\delta x_3-\delta x_1)- \mathrm{\Delta} x_{21}(\delta y_3-\delta y_1)\right]
\end{equation}
where $\mathrm{\Delta} x_{ab} = (x_{\mathrm{eq}})_{a} - (x_{\mathrm{eq}})_b$. Now, write each degree of freedom in terms of Fourier modes:
\begin{equation*}
    \delta \vec{x}_a = \frac{1}{\sqrt{N}}\sum_{\vec k}\delta x_{\vec k} e^{\mathrm{i}\vec k\cdot (\vec{x}_{eq})_a}.
\end{equation*}
where $N$ is the total number of particles. The Hamiltonian at lowest non-trivial order in fluctuations is then \begin{equation}
    H = \sum_k|\delta A_{\vec k}|^2 = \sum_{\vec k} \left(\begin{array}{ll} \delta x_{-\vec k} &\ \delta y_{-\vec k}\end{array}\right) \mathsf{M}_{\vec k} \left(\begin{array}{c} \delta x_{\vec k} \\ \delta y_{\vec k}\end{array}\right)
\end{equation}
where 
\begin{align}
    \delta A_{\vec k} &=\frac{1}{\sqrt{N}} \sum_{\text{triangle } s} \left[\delta y_{\vec k} \left(\mathrm{\Delta} x_{31} (\mathrm{e}^{\mathrm{i}\vec k \cdot \vec{x}_{2}}-\mathrm{e}^{\mathrm{i} \vec k\cdot \vec{x}_1})-\Delta x_{21}(\mathrm{e}^{\mathrm{i}\vec k \cdot  \vec{x}_3} - \mathrm{e}^{\mathrm{i}\vec k\cdot \vec{x}_1}) \right) \right. \notag \\ &\;\;\;\; \left.+\delta x_{\vec k} \left(\Delta y_{21}(\mathrm{e}^{\mathrm{i}\vec k \cdot  \vec{x}_{3}}-\mathrm{e}^{\mathrm{i} \vec k \cdot  \vec{x}_{1}}) - \Delta y_{31}(\mathrm{e}^{\mathrm{i}\vec k \cdot  \vec{x}_{2}}-\mathrm{e}^{\mathrm{i}\vec k \cdot  \vec{x}_{1}})\right)\right].
\label{eq:Area_fourier_mode}
\end{align}
The matrices $\mathsf{M}_{\vec k}$ are straightforwardly computing by collecting coefficients.  Hamilton's equations then imply that the normal modes propagate with a frequency \begin{equation}
    \omega_{\vec k} = \pm \sqrt{\det \mathsf{M}_{\vec k}},
\end{equation}
which are always real as the Hamiltonian has stable equilibrium.  Plugging in for the triangular lattice in \eqref{eq:xeq_triangular} we find \begin{equation}
    \omega(\vec k) = \pm \sqrt{48}\sin \frac{k_x}{2}\sin \frac{k_x+\sqrt{3}k_y}{4}\sin \frac{k_x-\sqrt{3}k_y}{4}.
\end{equation}

\begin{figure}
    \centering
    \begin{subfigure}[t]{0.3\textwidth} 
        \centering
        \includegraphics[width=\linewidth,keepaspectratio]{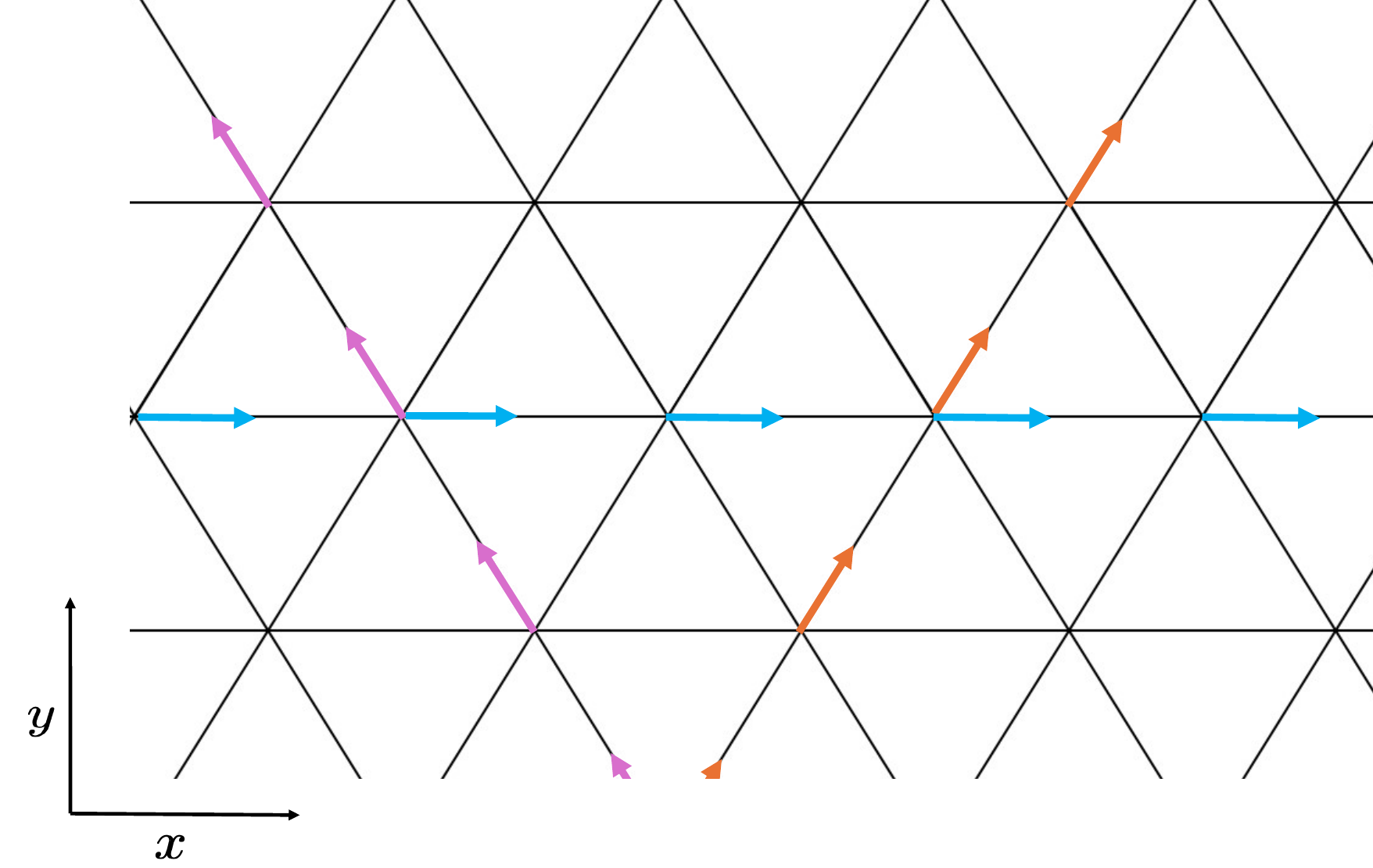} 
        \caption{}
        \label{fig:triangle_sub}
    \end{subfigure}
    \hfill
    \begin{subfigure}[t]{0.3\textwidth} 
        \centering
        \includegraphics[width=\linewidth, keepaspectratio]{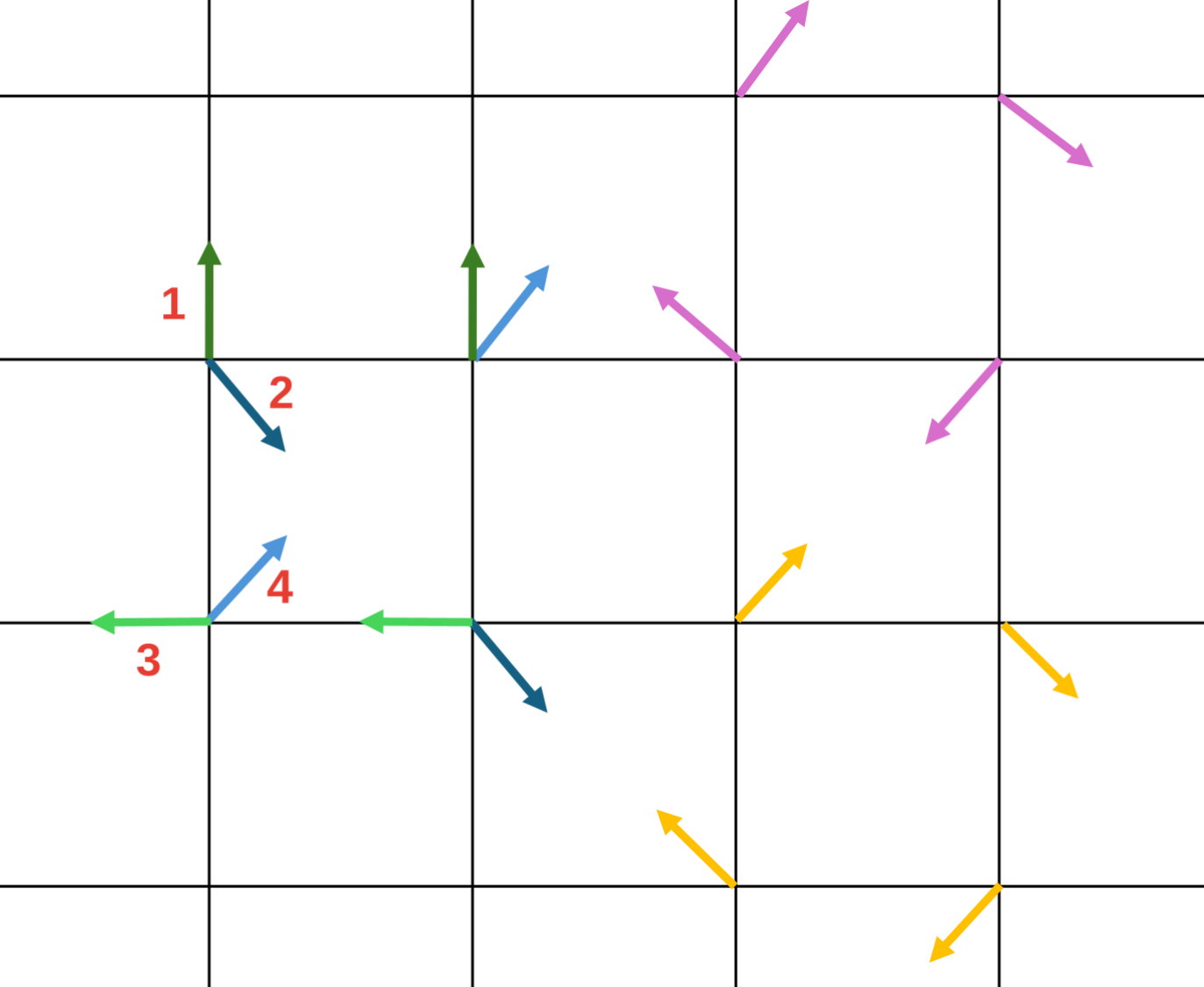}
        \caption{}
        \label{fig:square_sub}
    \end{subfigure}
    \hfill
    \begin{subfigure}[t]{0.3\textwidth} 
        \centering
        \includegraphics[width=\linewidth,keepaspectratio]{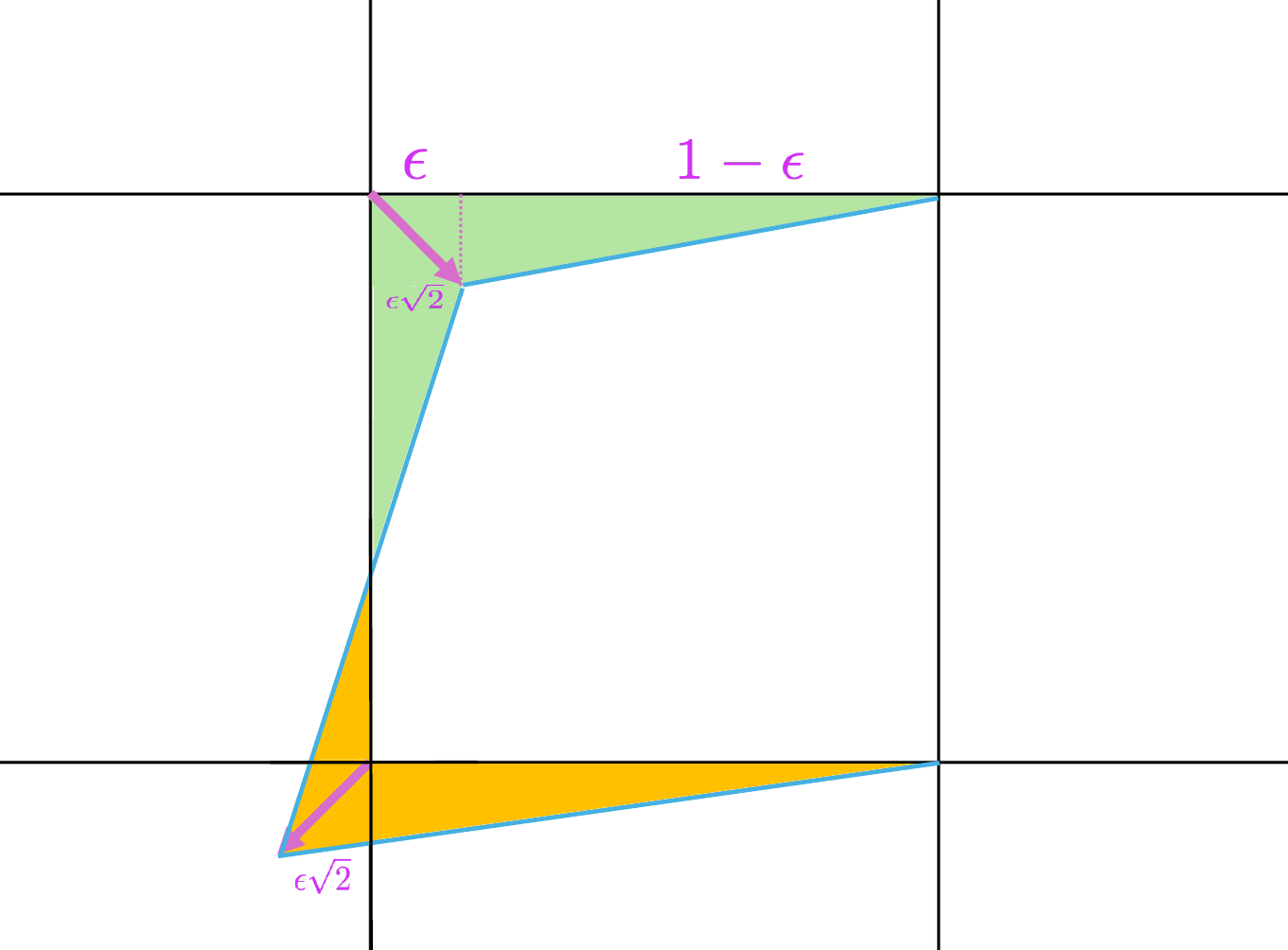} 
        \caption{}
        \label{fig:squarept2}
    \end{subfigure}
    
    \caption{Sub-system symmetries for the (a) triangular lattice and (b) square lattice. Notice in figure (b) of the square lattice, the yellow and purple arrows indicate two local transformations. The leftmost square represents the four subsystem symmetries one would naively guess at first glance. Figure (c) shows the change in area of the square adjacent to a square in which a local transformation (shown in figure (b)) is applied. The green shaded area is the area lost and the orange is the area gained. At first order in $\epsilon$, these areas are equal in magnitude, and the net change in area is $\Delta A \sim \epsilon^2$. Thus the transformation leaves both areas invariant and explains the proliferation of zero modes within linear response. }
    
    \label{fig:sidebyside}
\end{figure}

Note that $\omega=0$ when $k_x = 0$ or $k_x =\pm \sqrt{3}k_y$. These zero modes can be understood as follows. As shown in Figure \ref{fig:triangle_sub}, we can consider an arbitrarily large displacement of all particles in one row of the lattice, depicted with the blue arrows.  These displacements distort the lattice locally, but every triangle's area is unchanged as its base and height retain the same value.   In particular, we could apply such distortions independently to \emph{every} row of the lattice.   All of these displacements do not change the area of a triangle, and so we conclude that the space of configurations with $H=0$ is very high-dimensional, corresponding to not only translations/rotations/shears of the triangular lattice (which are guaranteed to leave energy unchanged by the symmetry group $\mathrm{SL}(2,\mathbb{R})\rtimes \mathbb{R}^2$), but the row-wise displacements described above and depicted in Figure \ref{fig:triangle_sub}.  This appears to correspond to a subsystem symmetry \cite{gromovfrachydro,Iaconis:2019hab} which explains the origin of the zero modes for any $k_x=0$, regardless of the value of $k_y$ (which corresponds to the relative strengths of the blue displacements on each row).  The same argument applies to the ``rows" of the lattice after rotating by $\pi/3$ and $2\pi/3$, explaining the other two families of zero modes.

  We do note, however, that these emergent subsystem symmetries of the triangular lattice are \emph{not} exact symmetries of the model, respected by the fully nonlinear dynamics.  Indeed, notice that if we try to apply two of these shifts along different lines in the lattice, we can change the area of a triangle where the lines intersect.  Instead, the emergence of this accidental subsystem symmetry is a peculiar artifact of the low-energy (linear response) regime of this specific model.

\subsection{More examples}
We now discuss a few additional examples of normal modes for different lattices.

A next example to discuss is square lattice, where each square unit cell represents a polygon $s$ in the sum \eqref{eq:Hamiltonian_triangle}. Repeating the calculation above we find that 
\begin{equation}
    \delta A_k = 2\mathrm{i}\mathrm{e}^{\mathrm{i}\vec k \cdot \vec x_{\text{cen}}} \left(\delta x \sin\left(\frac{k_x}{2}\right) \cos\left(\frac{k_y}{2}\right) + \delta y   \sin\left(\frac{k_y}{2}\right) \cos\left(\frac{k_x}{2}\right)\right)
\end{equation}
where $\vec x_{\text{cen}}$ denotes the equilibrium position of the center of the square.  A short calculation shows that
\begin{equation}
    \omega(\vec k) = 0.
\end{equation}

Let us understand how this happened.  On first glance, one might have expected to get a different answer:  $\omega\sim k_x k_y (k_x+k_y)(k_x-k_y)$, because like the triangular lattice, we should expect accidental subsystem symmetries associated with individually shearing each row/column/diagonal in the lattice: see Figure \ref{fig:square_sub}.  However, a more careful analysis, shown in the same Figure \ref{fig:square_sub}, reveals that for \emph{each unit cell} there is a local move that one can make which -- at linear order in $\delta \vec x$ -- does not change the area of a square.  One can rotate a single unit cell of the lattice by an infinitesimal angle $\epsilon$, while only changing the area of each unit cell by $\mathrm{O}(\epsilon^2)$.  This result is easiest to see pictorially, and is shown in Figure \ref{fig:squarept2}.   Consequently, the system cannot support propagating modes, and the dispersion relation collapses to zero for all excitations.

\begin{figure}[ht]
\centering

\begin{minipage}{0.48\textwidth}
    \centering
    \includegraphics[width=\linewidth]{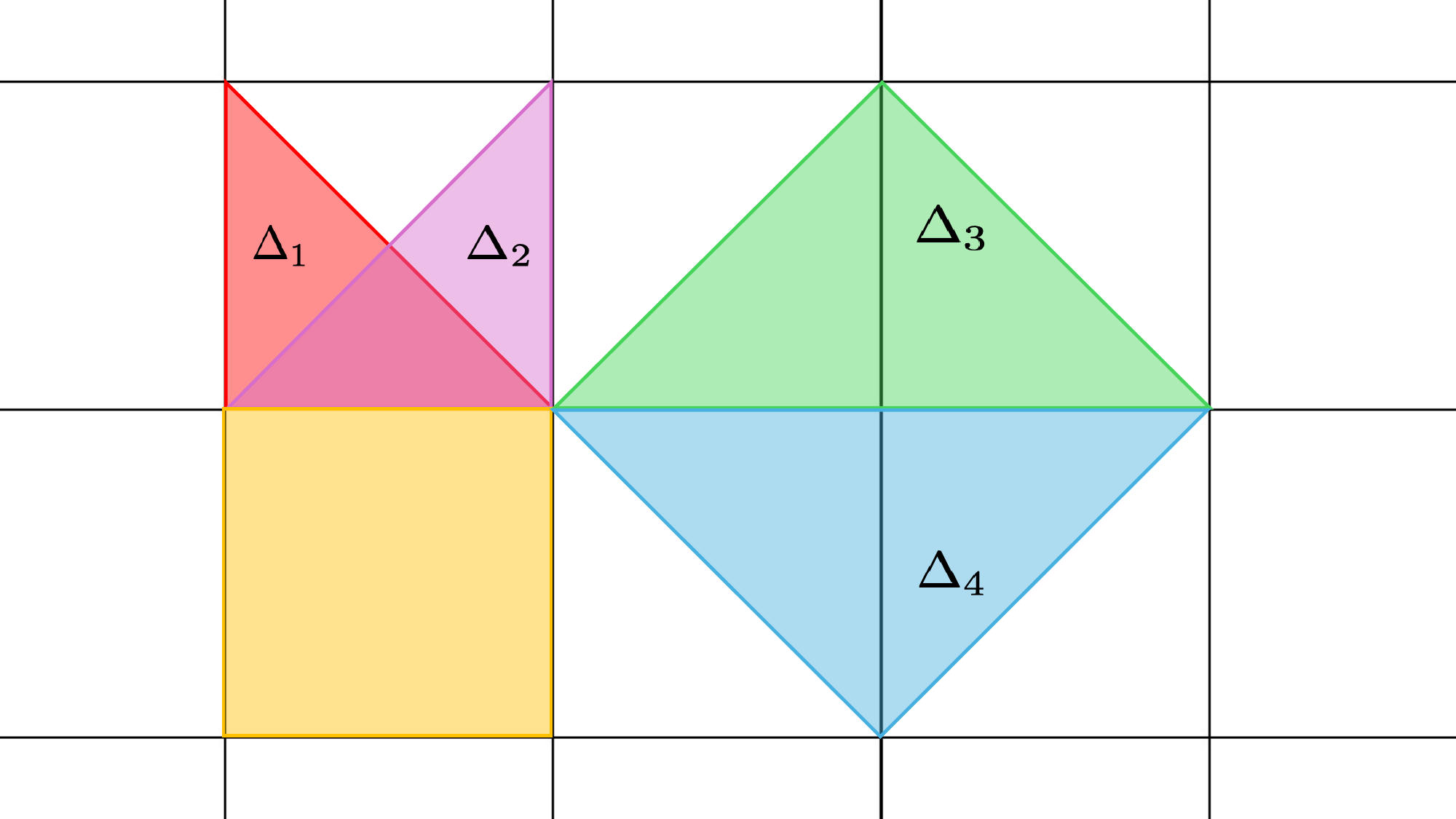}
\end{minipage}%
\hfill
\raisebox{0.05cm}{
\begin{minipage}{0.46\textwidth}
    \centering
    \small
    \renewcommand{\arraystretch}{1.3}
    \setlength{\tabcolsep}{4pt}
    \begin{tabular}{|l|l|}
    \hline
    \textbf{Hamiltonian} & \textbf{Dispersion Relation} \\ \hline
    $H_{\Delta_1}$ & $\omega^2 \approx \tfrac{1}{16}k_x^2k_y^2(k_x-k_y)^2$ \\ 
    $H_{\square}$ & $\omega^2 = 0$ \\ 
    $H_{\Delta_1} + H_{\Delta_2}$ & $\omega^2 \approx \tfrac{1}{4}k_x^2k_y^2(k_x^2+k_y^2)$ \\ 
    $H_{\Delta_3}$ & $\omega^2 \approx k_x^2(k_x+k_y)^2(k_y-k_x)^2$ \\ 
    $H_{\Delta_3} + H_{\Delta_4}$ & $\omega^2 \approx 4(k_x+k_y)^2(k_y-k_x)^2k_x^2$ \\ 
    $H_{\Delta_3} + H_{\Delta_1}$ & $\omega^2 \approx \tfrac{5}{16}k_x^2(k_x-k_y)^2(8k_xk_y+4k_x^2+5k_y^2)$ \\ 
    \hline
\end{tabular}
\end{minipage}
}

\caption{(a) Additional right triangle motifs that can be built on the square lattice.
(b) The $k_{x,y}\rightarrow 0$ limit of the dispersion relation coming from a Hamiltonian $H$.
The Hamiltonian $H_{\Delta_1}$ corresponds to the sum of $\Delta_1$-type triangles [from (a)] in each unit cell;
$H_{\square}$ corresponds to the square unit cell.}
\label{fig:Dispersion_plot_and_table}
\end{figure}

Now we will take the same square lattice model, but add additional triangles $s$ to the sum in \eqref{eq:Hamiltonian_triangle}.   Depending on which kinds of triangles $s$ are added to the sum, we can generate many different kinds of dispersion relations, which are illustrated in Figure \ref{fig:Dispersion_plot_and_table}.  

%Referencing figure \ref{fig:Dispersion_plot_and_table}.(a), then table \ref{fig:Dispersion_plot_and_table}.(b) gives the dispersion relations for Hamiltonians constructed by the following combinations of areas. 
\begin{figure}
    \centering
    \includegraphics[width=0.3\linewidth]{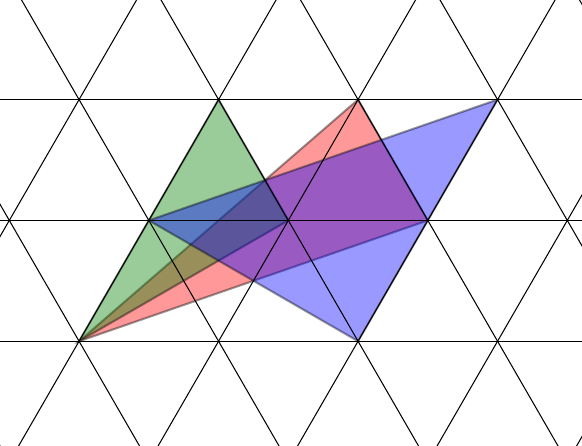}
    \caption{Configuration of triangles that enter Hamiltonian \eqref{eq:Hamiltonian_triangle} that has no subsystem symmetries.}
    \label{fig:nosubsym}
\end{figure}

It is also possible to remove all accidental subsystem symmetries by choosing sufficiently complicated polygons $s$ in the sum \eqref{eq:Hamiltonian_triangle}.  Figure \ref{fig:nosubsym} shows three types of triangles made from vertices on a triangular lattice.  When the $s$ in $H$ correspond to the areas of these triangles (translated by each unit cell), we find that (at small $k$)
\begin{equation}
    \omega^2 \approx 
116 k_x^6 + 348 k_x^5 k_y + 1885 k_x^4 k_y^2 - 754 k_x^3 k_y^3 - 1247 k_x^2 k_y^4 + 348 k_x k_y^5 + 348 k_y^6
\end{equation}
and indeed there is no accidental subsystem symmetry left.  

 $\omega \sim k^3$ scaling without any subsystem symmetries is indeed the most generic result one expects to find.  To derive this result and provide more intuition as to where it comes from, we now develop an effective field theory for the dynamics of these crystal lattices.

%\andy{Are the $x,y$ axes oriented the same way as in the figure?  It doesn't look like it to me.}

%\isabella{If you mean the sign on the $\delta x_i$ and $\delta y_i$ then no, I tried to make it as general as possible.}

\section{Effective field theory (non-dissipative)}\label{sec:eftnondiss}
To go beyond the linear response regime, it will be helpful to develop effective field theories (EFTs) for solid dynamics in the non-commutative plane.  We first do a ``warm-up" calculation where we present a non-dissipative EFT which captures the normal modes very close to the ground state.  A useful byproduct of the discussion is a better and more general understanding of the $\omega\sim k^3$ phonon dispersion relation.

To build an EFT it is usually helpful to work with Lagrangian, rather than Hamiltonian mechanics.  Writing  \begin{equation}
    L = \sum_{a=1}^n \frac{1}{2}\epsilon_{ij}\dot x_{ai}x_{aj} - H, \label{eq:discreteL}
\end{equation}
we can envision going to the continuum limit of an effective description, following \cite{Leutwyler:1996er,Dubovsky:2005xd,Nicolis:2013lma}, by replacing discrete particle label $a$ with two coordinates $(\sigma^1,\sigma^2)=\sigma^I$ which label the ``rest point" of that particle in the crystal.  Letting $X^i(\sigma,t)$ denote the physical position of that particle at time $t$, \eqref{eq:discreteL} implies a field theory Lagrangian of the form \begin{equation}
    \mathcal{L} = \frac{1}{2}\epsilon_{ij}\partial_t X_i X_j - \mathcal{H}(X_i, \partial_IX_j, \ldots). \label{eq:continuumL}
\end{equation}
Since we have only a single time-derivative in the first term of \eqref{eq:continuumL}, we expect higher time derivatives to be irrelevant and can neglect them.  In contrast, we must now deduce the constraints on $\mathcal{H}$ by finding invariant building blocks under the symmetry group.

Since $\mathrm{SL}(2,\mathbb{R})\rtimes\mathbb{R}^2$ acts affinely on $x_{ai}$, it should also act affinely on $X_i$.  The action must therefore be invariant under \begin{equation}
    X_i \rightarrow {M_i}^jX_j + \epsilon_i \label{eq:XSL2R}
\end{equation}
where $\epsilon_i$ is a constant vector and ${M_i}^j \in \mathrm{SL}(2,\mathbb{R})$ is an area-preserving coordinate transformation.  We saw in Section \ref{sec:hamiltoniansym} that $\epsilon_{ij}$ is the only invariant building block under $\mathrm{SL}(2,\mathbb{R})$, and as before, translation symmetry requires us to add at least one $\partial_J$ derivative.   We conclude that the simplest two invariant building blocks for our EFT are \begin{subequations}
    \begin{align} 
    \tau_{IJ} &= \partial_I X_i \epsilon^{ij}\partial_JX_j, \\
    \nu_{IJK} &= \partial_I X_i \epsilon^{ij}\partial_J\partial_KX_j
    \end{align}
\end{subequations}
And any Hamiltonian of the form $\mathcal{H}(\tau_{IJ}, \nu_{IJK}, \ldots)$ is consistent with all symmetry requirements imposed thus far.   We may (and will) consider further restrictions by imposing additional symmetries; alternatively, more microscopic details, such as crystal lattice symmetries in the ground state, could add further restrictions on $\mathcal{H}$.   For any such theory, as in \eqref{eq:symmetrygenerators}, the conserved quantities are \begin{equation}
    Q = \int \mathrm{d}^2\sigma \left[A^{ij}X_i X_j + B^iX_i\right] \label{eq:sec5conserved}
\end{equation}
for arbitrary vector $B^i$ and symmtric marix $A^{ij}$.

For simplicity in this section, we will focus on a ``homogeneous and isotropic crystal" which is invariant under $\mathrm{SO}(2)$ rotations of $\sigma_J$, along with translation, meaning that we must contract all $IJ$ derivative indices into each other in the Lagrangian.  The resulting EFT is then a bit simpler to analyze.  The most general invariant Lagrangian is  
\begin{equation}
    \mathcal{L} = \frac{1}{2}\epsilon_{ij}\partial_t X_i X_j - \frac{\alpha}{2}\tau_{IJ}\tau_{IJ}-\frac{\lambda}{2}(\tau_{IJ}\tau_{IJ})^2-\frac{\beta}{2}\nu_{IJK}\nu_{IJK} - \frac{\gamma}{2}\nu_{IJK}\nu_{JIK}- \cdots. 
\end{equation}
To understand the normal modes, we can now expand around the equilibrium configuration: \begin{equation}
    X_i \approx \delta_i^I(\sigma_I + \phi_I)  \label{eq:Xexpansion}
\end{equation}and keep only lowest non-trivial (quadratic) order terms in $\phi$, within the Lagrangian $\mathcal{L}$. We expand the relevant tensors as \begin{subequations}
    \begin{align}
        \tau_{IJ} &= \epsilon_{IJ}+\epsilon_{IK}\partial_J\phi_K + \epsilon_{KJ}\partial_I\phi_K+ \epsilon_{KL}\partial_I\phi_K\partial_J\phi_L,\\
        \nu_{IJK} &= \epsilon_{IL}\partial_J \partial_K \phi_L + \cdots,
    \end{align}
\end{subequations}
Substituting these into the Lagrangian and keeping only terms up to quadratic order, we obtain the leading-order effective theory,
\begin{align}
    \mathcal{L} &= \frac{1}{2}\epsilon_{IJ}\partial_t \phi_I \phi_J - (\alpha +12\lambda)(\partial_I\phi_I)^2 - 8\lambda (\partial_I\phi_I)^3 - 2(\alpha + 4\lambda)\epsilon_{IK}\epsilon_{MN}\partial_J\phi_K\partial_I\phi_M\partial_J\phi_N \notag \\
    &- \frac{\beta}{2}\partial_J \partial_I \phi_I\partial_J \partial_K \phi_K-\frac{\beta+\gamma}{2}\partial_I (\epsilon_{JK}\partial_J\phi_K)\partial_I  (\epsilon_{LM}\partial_L\phi_M) + \cdots. \label{eq:sec5H}
\end{align}
Looking for plane wave solutions to the Euler-Lagrange equations of motion for the theory, one finds after a short calculation that the normal modes of the EFT are of the form  $\phi_I \propto \mathrm{e}^{\mathrm{i}kx-\mathrm{i}\omega t}$, with

\begin{comment}
\begin{equation}
    \omega = \pm \sqrt{(\beta + \gamma)(4\alpha + \beta k^2)k^6}\approx \pm \sqrt{4\alpha(\beta+\gamma)} k^3.
\end{equation}
\end{comment}
\begin{equation}
    \omega \approx \pm \sqrt{(2(\alpha+12\lambda)k^2+\beta k^4)(\beta + \gamma)k^4} \approx \pm \sqrt{2(\alpha+12\lambda)(\beta + \gamma)}k^3
\label{eq:dispersion_relation_zeroT}
\end{equation}

We can understand the $k^3$ dispersion as follows: $\phi_x$ and $\phi_y$ are canonically conjugate variables.  For simplicity aligning the wave number in the $x$-direction, this means that a longitudinal mode $\phi_x$ must couple to the transverse one $\phi_y$.  The longitudinal mode enters $\mathcal{L}$ as $(k \phi_x)^2$, but the transverse mode enters $\mathcal{L}$ as $(k^2\phi_y)^2$.  The equations of motion multiply these two factors, giving the $k^3$ dispersion term.

The dispersion relation \eqref{eq:dispersion_relation_zeroT} was derived under full rotational symmetry $\mathrm{SO}(2)$, which reduces all possible quadratic gradient terms to two building blocks (divergence and curl). So, the coefficients that multiply these building blocks can only appear as sums, giving the two isotropic combinations $(\alpha + 12\lambda)$ and $(\beta + \gamma)$. If the microscopic theory instead prefers a crystal with a reduced point group symmetry, the possible coefficients of $\mathcal{L}$ can couple to all invariant tensors with $IJ\cdots$ indices, such as \begin{equation}
    \mathcal{L} = \frac{1}{2}\epsilon_{ij}\partial_t X_i X_j - \frac{1}{2}\alpha_{IJKL}\tau_{IJ}\tau_{KL}- \frac{1}{2}\beta_{IJKLMN}\nu_{IJK}\nu_{LMN}+\cdots.
\end{equation} Then the dispersion relation acquires explicit angular dependence, which is consistent with the many microscopic examples we saw in Section \ref{sec:lattices}.

\section{Effective theory of dissipative dynamics}\label{sec:eftdiss}

In this section, we now turn to discuss the dissipative EFT which describes the finite energy-density dynamics of most interest in the many-body system.

\subsection{Review of MSR EFT}
We first summarize how to construct a Hamiltonian effective field theory for dissipative dynamics relaxing to a thermal steady state.  We will simply state the prescription for how to do this without justification: the reader interested in more details should see \cite{huang2023generalized} for a pedagogical introduction.  A more sophisticated approach which can be used to derive similar results is the Schwinger-Keldysh path integral, reviewed in \cite{liu2018lectures}.

For simplicity, suppose for the moment that we have a discrete collection of variables $(q_1,\ldots, q_N)$, with a corresponding symplectic form $\omega_{ab}\mathrm{d}q_a\wedge \mathrm{d}q_b$ whose inverse $\omega^{-1} = V$.  A rather trivial Lagrangian whose Euler-Lagrange equations reproduce Hamilton's equations is \begin{equation}
    L_0 = \pi_a \dot q_a - T \pi_a V_{ab}\mu_b,
\end{equation}where we have defined \begin{equation}
    \mu_i = \frac{1}{T} \frac{\partial H}{\partial q_i}.
\end{equation}
Here $T$ denotes temperature, and its role will be elucidated shortly.  It turns out that this trivial Lagrangian can be extended non-trivially to include dissipative dynamics that obey a fluctuation-dissipation theorem -- this leads to the Martin-Siggia-Rose path integral \cite{PhysRevA.8.423} for stochastic dynamics.  Specifically, one modifies $L \rightarrow L_0+L_{\mathrm{diss}}$ where \begin{equation}
    L_{\mathrm{diss}} = \mathrm{i}T \pi_a \Sigma_{ab}\left(\pi_b-\mathrm{i}\mu_b\right).
\end{equation}
We notice that this Lagrangian transforms non-trivially under the transformations \begin{subequations} \label{eq:KMS}
    \begin{align}
        t &\rightarrow -t, \\
        q_a &\rightarrow \eta_aq_a, \\
        \pi_a &\rightarrow -\eta_a(\pi_a-\mathrm{i}\mu_a)
    \end{align}
\end{subequations}
where $\eta_a=\pm 1$ denotes whether the corresponding variable is even or odd under (generalized) time-reversal symmetry.  Indeed, \eqref{eq:KMS} can be thought of as a generalized time-reversal symmetry, or Kubo-Martin-Schwinger (KMS) symmetry, under which we must build an invariant Lagrangian.  To obtain a dissipative EFT therefore, we will build the most general Lagrangian invariant under KMS/time-reversal.  We may choose $L_0$ and $L_{\mathrm{diss}}$ defined above so long as \begin{subequations}
    \begin{align}
        \eta_a V_{ab}\eta_b &= -V_{ab}, \\
        \eta_a \Sigma_{ab}\eta_b &= \Sigma_{ab},
    \end{align}
\end{subequations}
namely $V$ is block off-diagonal and $\Sigma$ is block-diagonal between T-even/odd degrees of freedom.  

Turning to our particular construction, we could take $\eta_x=1$ and $\eta_y=-1$; namely the theory is invariant under a combination of parity and time-reversal symmetry.  Intuitively this makes sense, as this is the symmetry in e.g. the lowest Landau level.  To describe dissipative corrections to the Hamiltonian field theory of Section \ref{sec:lattices}, therefore, we would write \begin{equation}
    L = \pi_{ai}\dot x_{ai} - T\pi_{ai}\epsilon_{ij}\mu_{aj}+ \mathrm{i}T \Sigma_{ab}\pi_{ai}\left(\pi_{bi}-\mathrm{i}\mu_{bi}\right),
\end{equation}
with $\Sigma$ a symmetric and positive semi-definite matrix.

Let us now unpack the physics of this construction.  Evaluating Lagrange's equations for $\pi_{ai}$ we find \begin{equation}
    \dot x_{ai} = \epsilon_{ij}\frac{\partial H}{\partial x_{aj}} - \Sigma_{ab}\frac{\partial H}{\partial x_{bi}} + 2\mathrm{i}T\Sigma_{ab}\pi_b. \label{eq:MSRnaiveEOM}
\end{equation}
The first term above represents Hamilton's equations, while the second  term is a dissipative correction that tends to drive the system towards low energy states of $H$.  The final term above appears imaginary.  To understand the physics of this, one must interpret the Lagrangian above as defining a path integral.   One then notices that the final term in \eqref{eq:MSRnaiveEOM} corresponds to a Gaussian white noise field $2\mathrm{i}T\Sigma_{ab}\pi_{bi} \rightarrow \xi_{ai}$ obeying the fluctuation-dissipation theorem: \begin{equation}
    \langle \xi_{ai}(t)\xi_{bj}(s)\rangle = \delta(t-s)\cdot 2T\delta_{ij}\Sigma_{ab}.
\end{equation}
The second and third terms in \eqref{eq:MSRnaiveEOM} drive the system towards the statistical steady-state distribution \begin{equation}
    P_{\mathrm{ss}}(x_{ai}) = \mathrm{e}^{-H(x_{ai})/T}.
\end{equation}

Finally, in the presence of exact conserved quantities (often called ``strong symmetries"), one must demand that $L$ is invariant under not only KMS symmetry, but under the following transformation \cite{huang2023generalized} \begin{equation}
    \pi_{ai} \rightarrow \pi_{ai} + \frac{\partial F}{\partial x_{ai}},
\end{equation}
which realizes Noether's Theorem for this dissipative extension of Hamiltonian mechanics.  In other words, $\Sigma_{ab}$ must have null vectors in the direction $\partial_{ai}F$ for any conserved quantity $F$, and the Poisson bracket $\lbrace F,H\rbrace =0$.

It is straightforward to generalize this discussion to continuum mechanics.  One simply replaces the discrete variables $x_{ai}$ with continuous fields $\phi_i(x)$ and uses functional derivatives to define \begin{equation}
    \mu_i(x) = \frac{1}{T}\frac{\delta H}{\delta \phi_i(x)}.
\end{equation}
The rest of the MSR formalism goes through in a straightforward way.

% \begin{itemize}
%     \item We postulate a stationary state $\Phi$ which the system relaxes to, such that $P_{ss}=e^{-\Phi}$ is the stationary probability distribution. Additionally, we identify all relevant symmetries and specify a generalized time-reversal operation $gT$. 

%     \item We write the most general local Fokker-Planck generator $\hat{W}$ that conserves probability, annihilates $P_{ss}$, and is compatible with $gT$. This operator splits as $\hat{W}=\hat{W}_{even}+\hat{W}_{odd}$ where $\hat{W}_{even}$ is even under $gT$ symmetry, dissipative, and is fully fixed by $\Phi$. The $gT$-odd part is dissipationless and encodes non-reciprocal currents. Their interplay automatically implements a non-equilibrium fluctuation-dissipation relation. 

%     \item Organizing the terms allowed by symmetry in a derivative expansion, each labelled by its order in derivative and $gT$-parity, yields a controlled hiearchy of effective equations like in equilibrium EFTs. 
% \end{itemize}

% The Martin-Siggia-Rose construction then transforms $e^{-\hat{W}t}$ to a phase-space path integral with effective Hamiltonian $\mathcal{H}=-i\hat{W}$ inside the first order action which leaves us with the general Lagrangian form $\mathcal{L}= \pi_i q_i - H(\pi_i,q_i)$, where $\pi_i$ is the noise field. The $gT$ transformation plays an analogous role to the Kubo-Martin-Schwinger symmetry in thermal field theory. The effective Hamiltonian, $\mathcal{H}$, encodes the noise and inherits all symmetry (including $gT$ symmetry) and probabilistic constraints. 

\subsection{Application to quadrupole-conserving dynamics}
We are now ready to apply these methods to the problem of quadrupole-conserving dynamics.  Following the non-dissipative effective field theory of Section \ref{sec:eftnondiss}, and recalling that this EFT takes the form of a Hamiltonian EFT, we begin by quoting the leading Hamiltonian in terms of the fields $\phi_i$ (in this section, we will not bother to distinguish between $I$ and $i$ indices): \begin{equation}
    H = \int \mathrm{d}^2x \left[ \frac{a}{2}(\partial_i\phi_i)^2 + \frac{b}{2}\left(\partial_i \epsilon_{jk}\partial_j \phi_k\right)^2 + \frac{a^\prime}{3}(\partial_i\phi_i)^3 + \cdots \right]. \label{eq:sec6H}
\end{equation}
Compared to \eqref{eq:sec5H}, we have relabeled the coefficients in \eqref{eq:sec6H} for simplicity. The leading order hydrodynamics is given by a non-dissipative MSR Lagrangian \begin{equation}
    \mathcal{L}_0 = \pi_i \partial_t \phi_i + T\pi_i \epsilon_{ij}\mu_j.
\end{equation}
So far this is just a re-writing of the EFT of Section \ref{sec:eftnondiss}.

 Now we want to find the dissipative terms we can add. Recall that the symmetry group demands that the dissipative part be invariant under translations and $\mathrm{SL}(2,\R)$ transformations.  From \eqref{eq:sec5conserved} we deduce that the MSR EFT must be invariant under  \begin{equation}
     \pi_i \rightarrow \pi_i + B_i + A_{ij}x_j \label{eq:sec6pishift}
 \end{equation}
 for symmetric matrix $A$.  We observe that by construction, as \begin{equation}
    \mu_i = \beta \frac{\delta H}{\delta \phi_i} = -\partial_i \left[ a \partial_j \phi_j + a^\prime \left(\partial_j \phi_j\right)^2\right] + b \partial_l \partial_l \epsilon_{ji}\partial_j \epsilon_{mn}\partial_m \phi_n + \cdots ,
\end{equation}
$\mathcal{L}_0$ transforms by a total derivative under \eqref{eq:sec6pishift},  as \begin{equation}
    A_{ik}x_k \epsilon_{ij}\partial_j \left[a \partial_l \phi_l + \cdots \right]= \partial_j \left( A_{ik}x_k \epsilon_{ij}\left[a \partial_l \phi_l + \cdots \right]\right) - A_{ij}\epsilon_{ij} \left[a\partial_l\phi_l + \cdots \right]
\end{equation}
with the latter term vanishing since $A_{ij}$ is symmetric.

The leading order dissipative terms can be deduced by demanding that they are invariant under \eqref{eq:sec6pishift}.  In contrast to the subtle manipulations of the previous paragraph, enforcing that the dissipative terms are invariant under \eqref{eq:sec6pishift} is relatively direct: \begin{equation}
    \mathcal{L}_{\mathrm{diss}} = \mathrm{i} T\kappa \epsilon_{ij}\partial_i \pi_j  \epsilon_{kl}\partial_k \left(\pi_l - \mathrm{i}\mu_l\right) + \mathrm{i} T\lambda_{ijk,lmn} \partial_i\partial_j \pi_k \partial_l\partial_m \left(\pi_n - \mathrm{i}\mu_n\right) + \cdots 
\end{equation}
where $\kappa \ge 0$ and $\lambda$ is a positive semi-definite tensor.

We can now deduce the quasinormal modes predicted by hydrodynamics. If $\vec{k}=(k,0)$; then the linearized equations of motion are
\begin{subequations}
    \begin{align}
        -\mathrm{i} \omega   \phi_x &\approx  -\lambda_{xxx,xxx}ak^6\phi_x -bk^4 \phi_y \\
        -\mathrm{i}\omega \phi_y &\approx a k^2 \phi_x - \kappa b k^6\phi_y
    \end{align}
\end{subequations}
We have kept only the leading-order terms in the above equation.
Solving for $\omega$ and expanding its exact roots in the long-wavelength limit $k\to 0$ and retaining only the leading dissipative contribution from the $\kappa$ term (order $k^{6}$) we obtain
\begin{equation}
\omega_{\pm}(k) \approx \pm \sqrt{ab}k^3 - \frac{\mathrm{i}}{2}k^6(b\kappa + a\lambda_{xxx,xxx})\label{eq:disslinear}
\end{equation}
So we have a pair of propagating modes with the familiar $k^3$ dispersion and sub-leading damping $\sim k^6$. 

In equilibrium, solids do not exist in the true thermodynamic limit in two spatial dimensions \cite{halperin}.  In other words, in the true thermodynamic limit, $\phi_i$ cannot be degrees of freedom in the genuine EFT, because they are Goldstone bosons but their corresponding symmetry (translation) cannot be broken spontaneously.  Therefore, on the very longest length scales -- well beyond accesibility in numerical simulations --  we expect that the effective field theory of this model looks rather different.   Without any Goldstone bosons $\phi_i$, the only slow degree of freedom is the charge density $\rho$.  The analysis of \cite{gromovfrachydro,Glorioso:2023chm} then implies that the resulting effective theory would be \begin{equation}
\label{eq:densitymodes}
    \partial_t \rho = D \left(\nabla^2\right)^3 \rho.
\end{equation}

\subsection{Instability of linear response hydrodynamics}
It is useful to do a scaling analysis on the MSR action 
\begin{equation}
    S = \int \mathrm{d}t\mathrm{d}x\mathrm{d}y \left(\pi_i\partial_t \phi_i + \pi_i\mu_j\epsilon_{ij} + \mathrm{i} \kappa \epsilon_{ij}\partial_i\pi_j \epsilon_{kl}\partial_k(\pi_l - \mathrm{i}\mu_l) + \mathrm{i} \lambda_{ijk,lmn} \partial_i\partial_j \pi_k \partial_l\partial_m \left(\pi_n - \mathrm{i}\mu_n\right) +\cdots \right). \label{eq:sec63S}
\end{equation}
Although it is not completely obvious due to the clear anisotropy between transverse and longitudinal modes, we go ahead and assume that \begin{subequations}
    \begin{align}
        [\partial_x]=[\partial_y] &= 1, \\
        [\partial_t] &= z.
    \end{align}
\end{subequations}
As we want the steady state $H$ to be dimensionless, this requires \begin{equation}
    [\phi_i] = 0.
\end{equation}
Following \cite{Glorioso:2021bif}, we allow for quadratic dissipationless terms to be relevant, instead fixing $z$ by demanding that the leading order dissipative correction is marginal.  KMS symmetry requires that \begin{equation}
    [\pi_i] = 2[\partial_i ] + [\phi_i] = 2.
\end{equation}
Then, comparing the dimensions of the first and fourth terms in \eqref{eq:sec63S}, we deduce that $z=6$ in agreement with what we found previously.   On the other hand, the dimension of $(\epsilon_{ij}\partial_i \pi_j)^2$ is smaller than the fourth term, and suggests $z=4$.  In either case, we will find relevant nonlinearities.
%Although the first dissipative term scales as $z=4$, in the absence of Goldstone fields $\phi_i$, we obtain dissipation that scales as $z=6$. Since the full $\phi_i$ theory reduces to \eqref{eq:densitymodes} when the Goldstone modes are not slow, it is natural to adopt $z=6$ as the scaling fixed point. The naive $z=4$ prediction from the first dissipative term actually becomes more irrelevant than the $z=6$ that the symmetry actually enforces. In other words, the symmetry has promoted the $k^6$ structure to be the leading dissipative invariant. This is consistent with what we found in \eqref{eq:disslinear}. 

  Crucially, we recall that $H$ is not a quadratic function: the $a^\prime$-term in \eqref{eq:sec6H} will contribute to the non-dissipative terms in the equations of motion, and is potentially relevant.  Such relevant nonlinearities also arise in dipole and momentum conserving fracton hydrodynamics \cite{Glorioso:2021bif,Glorioso:2023chm} as well as in the Navier-Stokes equations in one spatial dimension \cite{KPZ}.   We find that (taking $z=4$) \begin{equation}
    [\pi_y] + [\mu_x]_{a^\prime} = 2 + [\partial_x (\partial_x \phi_x)^2] = 5 < [\pi_i \partial_t \phi_i ] = 6.
\end{equation} Whether we take $z=6$ or $z=4$, we deduce that the $a^\prime$ nonlinearity is relevant, and following the previous examples mentioned, that the ultimate value of $z$ will be limited not by the naive dissipative terms, but by this nonlinearity.  Demanding that this nonlinearity is marginal and that the $\kappa$ term is consequently irrelevant, at tree-level we find the crude estimate \begin{equation}
    z = [\partial_x (\partial_x \phi_x)^2]= 3. \label{eq:z3}
\end{equation}

\section{Numerical simulations}\label{sec:numerics}
In this section, we describe numerical simulations of $N^2 \sim 10^4$ particles in a triangular lattice.  For simplicity we take periodic  boundary conditions. Formally this renders the definitions of dipole and quadrupole moments ill-defined; practically, the dynamics is still generated by the same Hamiltonians that we described previously, and so we should still expect to see the same universality class of dynamics arise.

More precisely, we consider a system whose ground state represents a right triangular lattice with each unit cell having $a=1$ side length. \begin{equation}
    H = \sum_{s=1}^{T} H_s
\end{equation}
where  $T=2(N-1)^2$ is the total number of small triangles in the lattice, and \begin{equation}
    H_s = \left(A_s - \frac{1}{2}\right)^2.
\end{equation}
$A_s$ denotes the area of the $T$ fundamental triangles that make up the lattice.
Indeed since each unit cell has area $1/2$, this Hamiltonian has the desired ground state.

Following the discussion in Section \ref{sec:lattices}, we calculate the linear response dispersion relation for this Hamiltonian to be
\begin{equation}
\label{eq:full_disp_triangle}
    \omega^2 = 4 \sin^2\left(\frac{k_y}{2}\right)\left(\cos\left(k_x-\frac{k_y}{2}\right)-\cos\left(\frac{k_y}{2}\right)\right)^2
\end{equation}
And for small $k$ we find that
\begin{equation}
\label{eq:disp_relation_triangle}
    \omega^2 \approx \frac{1}{4}k_x^2k_y^2(k_x-k_y)^2.
\end{equation}
We have numerically verified the dispersion relation in \eqref{eq:full_disp_triangle} by evolving data with normal mode initial conditions on a triangular lattice with periodic boundary conditions for $10^2$ particles and extracting the oscillation frequency from the time series. For each allowed lattice wave vector $(k_x,k_y)$ we initialize the displacement in that single mode and confirm that the measured $\omega^2$ agrees with \eqref{eq:full_disp_triangle} to within numerical accuracy. 

\subsection{Trotter method}
 Solving the full set of Hamilton equations would require solving $2N^2$ coupled differential equations.  Because our universality class is sensitive to the quadrupole-conservation laws, we would like a numerical time evolution algorithm that manifestly respects these conservation laws. Since an analytic solution (\ref{single_triangle_eq}) for the dynamics of a single triangle exists, we adopt a first-order Trotter splitting algorithm \cite{Trotter1959} to numerically simulate the Hamiltonian dynamics \cite{forest1990}.  Recall that in Hamiltonian mechanics, the time evolution of an arbitrary observable $A$ is generated by $A(t) := \exp[-\mathrm{ad}_H \cdot t] A$, where $\mathrm{ad}_H A := \lbrace H,A\rbrace$ with $H$ the Poisson bracket.  This is the classical analogue of the Heisenberg picture in quantum mechanics. For a small time step $\tau$, 
\begin{equation}
\mathrm{e}^{-\tau \cdot \mathrm{ad}_H} = \mathrm{e}^{-\tau \cdot ( \mathrm{ad}_{H_1} + \cdots + \mathrm{ad}_{H_T})} = \prod_{j=1}^T \mathrm{e}^{-\tau \cdot \mathrm{ad}_{H_j}} + \mathrm{O}\left(\tau^2\right). \label{eq:trotter}
\end{equation} 
 If $\tau$ is sufficiently small, then the error in this approximation should be negligible.  We also note that the time step $\tau$ does not a priori need to rapidly vanish with increasing system size: due to spatial locality \cite{PhysRevX.11.011020}, the Trotter errors do not scale with system size when studying local observables at short times.  We postulate that the hydrodynamic behavior can be accurately captured out to longer time; see also \cite{pollmannhydro}.

To do numerical simulations, we pick a random permutation of the $H_j$ and apply the Trotterized time evolution for a small time step $\tau=10^{-2}$, as in \eqref{eq:trotter}.  The same ordering is applied in each time step. We have chosen to use this algorithm because -- up to numerical precision -- the Trotterization \eqref{eq:trotter} \emph{exactly} conserves the dipole and quadrupole moments.  After all, we may use the exact evolution of a single triangle, given in \eqref{single_triangle_eq}, to all orders in $\tau$.  To check the accuracy of the single triangle evolution method, we compared its normal mode spectrum with that obtained from the direct integration of the $2N^2$ coupled equations of motion on smaller system sizes $N$. The frequencies agree within numerical precision.  In what follows, we used initial conditions \begin{equation}
    (x_{n_1n_2}(0),y_{n_1n_2}(0)) = \left(n_1a+\frac{n_2 a}{2} + \Delta_{i,x}a,  \frac{\sqrt{3}}{2}n_2a+\Delta_{i,y}a\right) \label{eq:Delta}
\end{equation}
where $\Delta_{n_1n_2,x/y}$ are uniform random variables in $[-\Delta,\Delta]$.
Here $0\le n_1,n_2 < N$ denote the coordinates on the lattice of each particle.  To account for periodic boundary conditions, when evaluating the areas of triangles at the boundaries, the coordinates are appropriately shifted.  We run each simulation to final time $t_{\mathrm{max}}=400$.

% Unless stated otherwise, the following parameters were used in the simulation:
% \begin{itemize}
%     \item Lattice spacing set to $a=1$.
%     \item Time step set to $\Delta t = 0.01$.
%     \item Total number of steps, $N_{step} = 4\times10^4$.
%     \item Initial conditions: each particle receives an independent random shift
%         \((\delta x_{i},\delta y_{i}) \in [-\Delta a,\Delta a]\) with \(\Delta = 0.3\),
%         eliminating unwanted subsystem symmetries.
% \end{itemize}

\subsection{Results}
To showcase the numerical results, it is natural to study correlation functions of hydrodynamic variables.   We will focus on the density \begin{equation}
    \rho(\vec x, t) = \sum_{j=1}^{T}\delta (\vec x - \vec x_j(t)).
\end{equation}
It is convenient to instead perform a spatial Fourier transform:
\begin{equation}
    \rho(\vec{k},t) = \sum_{j=1}^T \mathrm{e}^{\mathrm{i}\vec{k}\cdot \vec{x}_j(t)}
\end{equation}
To measure the dynamical critical exponent $z$, which we predicted to be $z\approx 3$ after accounting for relevant nonlinearities in \eqref{eq:z3}, we numerically compute the density two-point function
\begin{equation}
\label{eq:correlator}
    C(\vec k,t) = \langle \rho(-\vec k,t)\rho(\vec k,0)\rangle.
\end{equation}
 High-wave number modes decay faster than low-wave number modes, causing the system to thermalize quickly at high momentum scales and gradually transition to a state dominated by low-momentum fluctuations.

\begin{figure}
\centering
\begin{subfigure}[t]{0.48\textwidth}
    \centering
    \includegraphics[width=10cm,keepaspectratio]{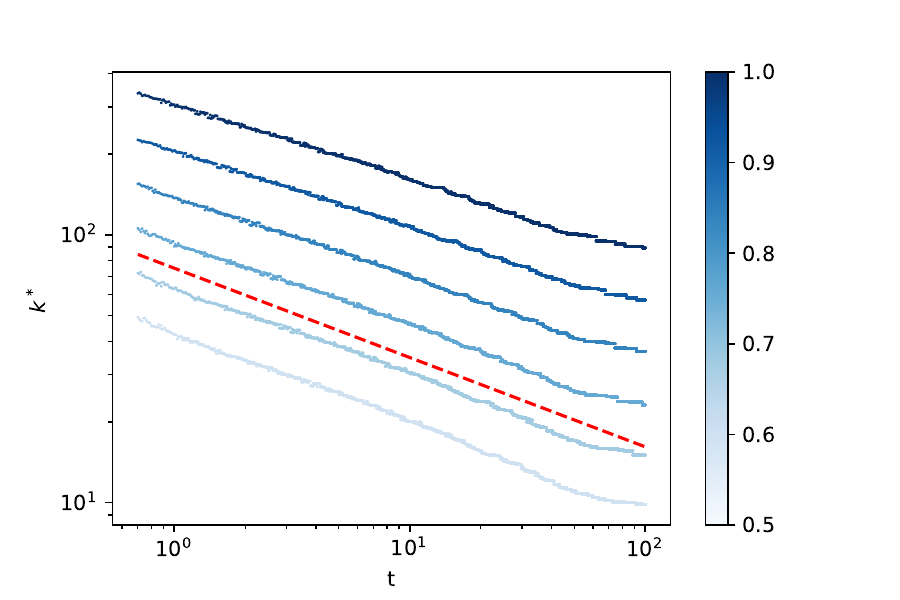}
    %\caption{The above image shows the logarithmic plot of $k^*(t)$ vs. $t$ for varying energy density. The red line is the line for $k^*(t) \sim t^{-1/3}$ or $z=3$.}
\end{subfigure}
\hfill
\begin{subfigure}[t]{0.48\textwidth}
    \centering
     \includegraphics[width=9cm,keepaspectratio]{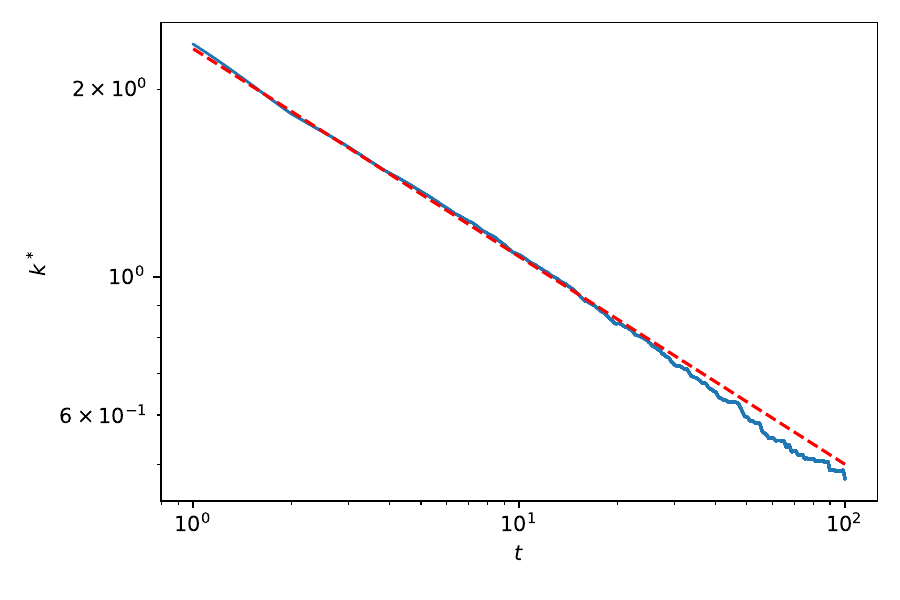}
    %\caption{Logarithmic plot of $k*(t)$ vs. $t$ for a fixed energy density with maximum fluctuations $0.25a$ where $a$ is the lattice spacing. The simulations were done on a $200\times 200 $ particle triangular lattice with periodic boundary conditions averaged over 30 runs. The red line again represents $k*(t) \propto t^{-1/3}$}
\end{subfigure}
\caption{Log-log plots of the characteristic wavenumber $k^*(t)$ vs. $t$ on a $150\times 150$ particle triangular lattice with periodic boundary conditions. Each point represents an average over 30 independent simulations. Figure (a) shows the scaling behavior at various energy densities, demonstrating that the dynamical law is insensitive to the choice of initial energy.  Figure (b) shows data for fixed maximum energy fluctuations set to $0.3$; the red dashed line shows $z=3$ scaling. Although the numerics were run for $T= 400$, we only used data up until $T/4$. This is because at later times, the autocorrelation function samples from far fewer data points, making the result sensitive to noise.}
\label{fig:z_plots}
\end{figure}

  In order to quantify this decay rate of high-wave number modes we define $k^*(t)$ via \begin{equation}
      C(k^*(t),t) = 0.7 \max_k C(k,t).
  \end{equation} The prefactor of 0.7 is not crucial, but accounts for the fact that even at early times there is a small preference for stronger correlations at small $k$.  We estimate that the correlations at wave number $k$ should decay exponentially quickly: $C(k,t) \sim \exp[-a\cdot k^z t]$ for some phenomenological constant $a$.  This implies that \begin{equation}
      k^*(t) \sim t^{-1/z}.
  \end{equation} 
Computing $k^*(t)$ using simulations of $\sim 10^4$ particles, we find that all simulations are consistent with $z\approx 3$.  Figure \ref{fig:z_plots} shows that this result is robust to the initial energy density of the model, which we adjusted by modifying the parameter $\Delta$ described below \eqref{eq:Delta}, over the numerically accessible time scales, with over an order of magnitude of $z=3$ scaling visible.

\begin{figure}
\centering
\begin{subfigure}[t]{0.48\textwidth}
    \centering
    \includegraphics[width=9cm,keepaspectratio]{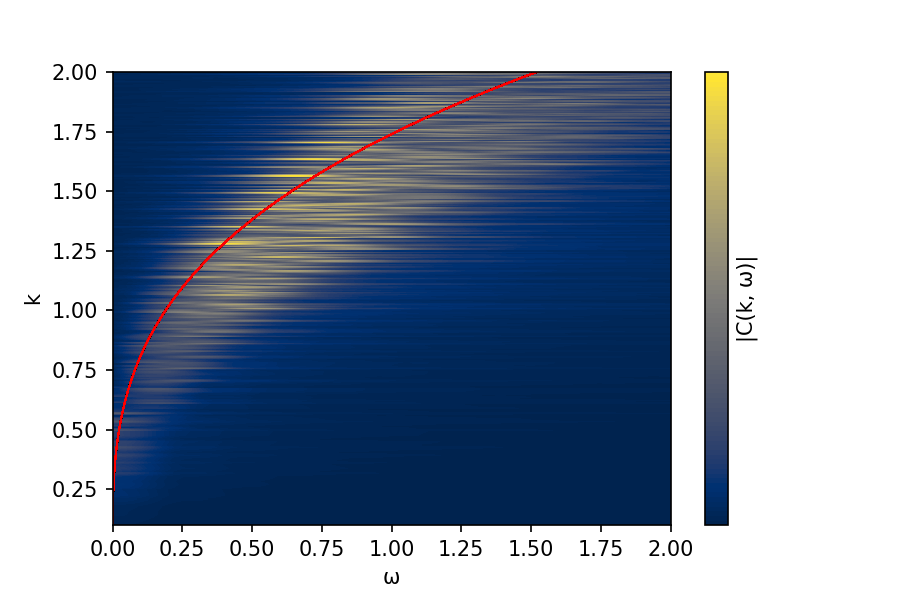}
 
\end{subfigure}
\hfill
\begin{subfigure}[t]{0.48\textwidth}
    \centering
     \includegraphics[width=9cm,keepaspectratio]{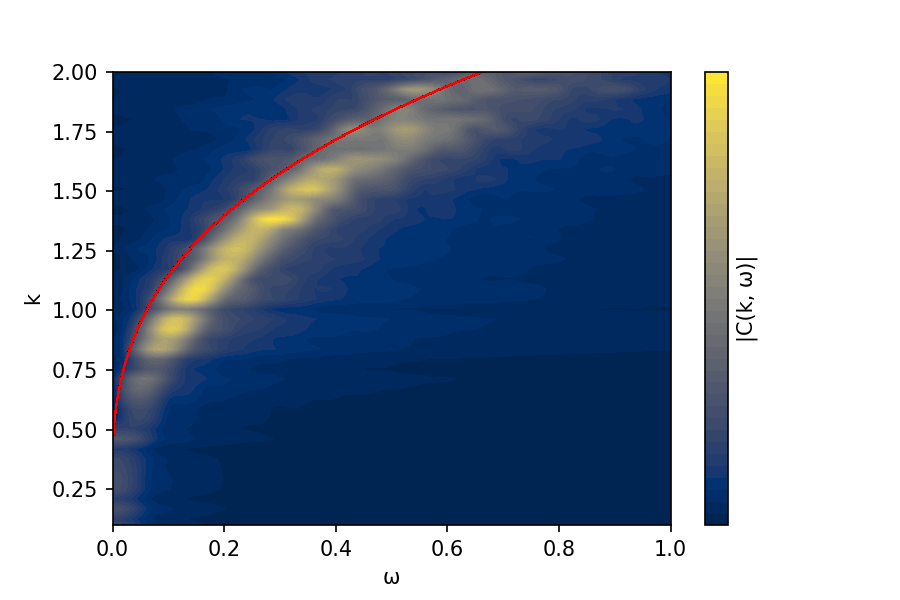}
\end{subfigure}
\caption{Color map of the magnitude of the temporal Fourier transform of the normalized density–density correlator, $C(|\mathbf{k}|,\omega)$. In Figure (a), for each frequency $\omega$ we average over all wave-vectors that share the same magnitude $|\mathbf{k}|$ (radial $k$–shells). The peak in the correlation function is relatively close to the cubic dispersion $\omega=\alpha k^{3}$ with $\alpha=1/\sqrt{32}$, in agreement with the analytic result detailed in Section 3, averaged over angles. In Figure (b), no radial averaging is performed: we show a single $(k_x,\omega)$ slice of the normalized density-density correlator with $k_y = 2\pi/15$ fixed. For both figures, the red line is the exact analytic prediction calculated in \eqref{eq:disp_relation_triangle}. Results are obtained from simulations on a $150\times150$ triangular lattice and averaged over 20 independent realizations.}
\label{fig:temporal_FT}
\end{figure}

Figure \ref{fig:temporal_FT} shows a two-dimensional plot of the normalized temporal Fourier transform of the correlator function $|C(k,\omega)|$ given in \eqref{eq:correlator}.  In Figure \ref{fig:temporal_FT}(a) we averaged over all radial $k$-shells and found that the dispersion relation is consistent with $\mathrm{Re}(\omega)\sim k^3$, in agreement with the linear response hydrodynamics.  Indeed, following \cite{Glorioso:2021bif}, we did not expect the real part of the dispersion relation to pick up a new scaling exponent.  The extremely broad nature of $|C(k,\omega)|$ is consistent with the ``sound mode" of the hydrodynamic theory being nearly over-damped, with $\omega \sim k^3-\mathrm{i}k^3$ predicted using the tree-level effective field theory.

Additionally, we found the strong angular dependence in $C(k,\omega)$ which is expected, recalling from Section \ref{sec:lattices} that the triangular lattice should have emergent subsystem symmetries of the hydrodynamics within linear response.

\section{Outlook}
In this paper we have described the ``fracton hydrodynamics" of a crystal in the non-commutative plane undergoing dynamics generated by a quadrupole-conserving Hamiltonian. The symmetry algebra of this model is the largest rotationally-invariant multipole algebra in the non-commutative plane that is finite-dimensional.  We saw that the resulting dynamics has a rather interesting interpretation: it is associated with dynamics that preserves the local area of shapes, so it is intrinsically associated with three-particle terms in the Hamiltonian, as in previous constructions of quadrupole-conserving Hamiltonian models \cite{osborne}. 

Neglecting the effects of energy conservation, we predicted analytically and observed numerically a clear breakdown of hydrodynamic linear response theory and a non-trivial dynamical universality class governing the long-wavelength physics of the crystal lattice.   

Unfortunately, we cannot think of a practical experimental realization for this universality class.  It was recently proposed that certain Galilean-invariant quantum Hall crystals can realize a ``fracton hydrodynamic" universality class with fewer symmetries \cite{Huang:2024qhx}.  Superfluid vortex crystals \cite{Du:2022xys,Glodkowski:2025krf} may also behave similarly, although one must account for the additional spontaneously broken U(1) symmetry, as well as the fact that dissipative dynamics could break multipolar conservation laws \cite{ambegaokar,qi2023kinematicallyconstrainedvortexdynamics}.

Nevertheless, if one can find a natural home for area-preserving dynamics in a quantum Hall or vortex fluid settings -- or engineer such dynamics in synthetic matter -- it may be possible to realize some of the unusual dynamical systems we have introduced in this work.   Given the large challenge of finding a system that has intrinsically three-body interactions in a quantum Hall regime or superfluid vortex crystals (let alone demanding quadrupole conservation), we might look instead to engineer our model in a toy classical system. To study dynamics in a non-commutative plane, it is possible to both rotate a system and place it in a harmonic trap, designed so that the centrifugal and harmonic restoring forces cancel (in the rotating non-inertial frame); the residual Coriolis force plays the role of a magnetic field in the (quantum Hall) regime.  To realize a system whose potential energy depends only on the area of a shape, we can envision filling the voids in a two-dimensional lattice of particles (recall the models in Section \ref{sec:lattices}) with a nearly incompressible fluid, so that changing the area of any triangle drastically changes the energy.  As in Section \ref{sec:lattices}, the particles should be located at the vertices of our engineered lattice.   We do not know, however, how to avoid two-body interactions that might naturally arise from whatever material contains the fluid within each triangle in the lattice; e.g. any semi-flexible rod that joins two particles.  If this issue can be solved, then our suggestion could lead to an experimental realization of this new dynamical universality class.

\section*{Acknowledgements}
AL thanks Andrey Gromov for an early collaboration related to this work at the Simons Center for Geometry and Physics. We also thank Xiaoyang Huang, Des Johnston, and Umang Mehta for useful discussions. This work was supported by the National Science Foundation through CAREER Grant DMR-2145544.

\section*{Data availability}

The data that support the findings of this article are openly available \cite{ZaneLucas2025code, Zane2025data}.

\bibliography{references}

\end{document}